\documentclass[10pt,aps,prd,twocolumn,superscriptaddress,showpacs,nofootinbib,noeprint]{revtex4-1}

\usepackage[utf8]{inputenc}

\usepackage{diagbox}

\usepackage{mathtools}
\usepackage{amsfonts}
\usepackage{mathrsfs}
\usepackage{bbm}
\usepackage{slashed}

\usepackage{graphicx}
\usepackage{color}
\usepackage{array}
\usepackage{esint}
\usepackage{placeins}
\usepackage{booktabs}
\usepackage{makecell}
\usepackage{epstopdf}
\usepackage[caption=false]{subfig}

\usepackage{xspace}
\usepackage{siunitx}
\usepackage{hyperref}
\usepackage[nameinlink]{cleveref}
\usepackage{appendix}

\usepackage{xifthen}
\usepackage{xcolor}
\hypersetup{
	colorlinks,
	linkcolor={red!75!black},
	citecolor={blue!75!black},
	urlcolor={blue!75!black}
}

\usepackage{comment}
\usepackage{dsfont}
\usepackage{tensor}
\usepackage{tabularx}

\begin{document}


\title{Anisotropic quark propagation and Zeeman effect in an external magnetic field}


\author{Minghui Ding}
\email{mhding@nju.edu.cn}
\affiliation{School of Physics, Nanjing University, Nanjing, Jiangsu 210093, China}
\affiliation{Institute for Nonperturbative Physics, Nanjing University, Nanjing, Jiangsu 210093, China}
\affiliation{Helmholtz-Zentrum Dresden-Rossendorf, Bautzner Landstra{\ss}e 400,
01328 Dresden, Germany}

\author{Fei Gao}
\email{fei.gao@bit.edu.cn}
\affiliation{School of Physics, Beijing Institute of Technology, 100081 Beijing, China}

\author{Sebastian \,M.~Schmidt}
\email{s.schmidt@hzdr.de}
\affiliation{Helmholtz-Zentrum Dresden-Rossendorf, Bautzner Landstra{\ss}e 400,
01328 Dresden, Germany}
\affiliation{Technische Universit\"at Dresden, 01062 Dresden, Germany}



\date{\today}

\begin{abstract}

We investigated the impact of a constant external magnetic field on the dressed propagators of up-, down-, and strange quarks. In the weak field limit, we derive a general momentum-space representation for the propagator and numerically solve the corresponding gap equation. Our analysis reveals that the vector term of the propagator can be decomposed into components parallel and perpendicular to the magnetic field, resulting in anisotropic effective masses, with the transverse mass consistently exceeding the longitudinal mass. This mass disparity exhibits a power law dependence on the magnetic field strength and is less pronounced for the strange quark compared to up and down quarks. Additionally, the magnetic field induces axial-vector and tensor terms, highlighting the Zeeman effect resulting from quark interactions with the magnetic field. These findings have important implications for (inverse) magnetic catalysis, and phenomena such as vector meson and pion condensations.

\end{abstract}


\maketitle



%

\section{Introduction}

The physics of strong electromagnetic fields offers a wealth of phenomena across various domains. In quantum electrodynamics (QED), the advent of high-intensity laser facilities, particularly following the invention of chirped pulse amplification, opens avenues for exploring non-perturbative effects in the vacuum, such as Schwinger electron-positron pair production and vacuum birefringence~\cite{Schutzhold:2008pz,Ahmadiniaz:2024xob}. These investigations aim to elucidate the fundamental nature of vacuum in strong electromagnetic fields. In quantum chromodynamics (QCD), the study of strongly interacting matter in the presence of intense electromagnetic fields enhances our understanding of the strong interaction and QCD matter behavior in extreme environments. Recent interest has surged in studying matter under external magnetic fields, as evidenced by various reviews, e.g.~\cite{Andersen:2014xxa,Miransky:2015ava,Fukushima:2018grm,Hattori:2023egw}. This focus is driven by the recognition that strong magnetic fields can exist in various astrophysical and experimental environments, such as the early universe~\cite{Grasso:2000wj}, compact stars (notably magnetars)~\cite{Duncan:1992hi}, and non-central heavy-ion collisions particularly ultra-peripheral collisions~\cite{Bertulani:2005ru} at facilities such as the Relativistic Heavy Ion Collider (RHIC) and the Large Hadron Collider (LHC)~\cite{Skokov:2009qp}. Despite their transient nature, the magnetic fields generated at these facilities can reach extraordinary strengths, thousands of times greater than those on magnetar surfaces. At RHIC, for instance, magnetic fields can reach approximately $eB \sim m_\pi^2$,  about $10^{18}$ Gauss, while at the LHC, they can intensify to $ eB \sim 10 \, m_\pi^2 $, approximately $ 10^{19}$  Gauss~\cite{Skokov:2009qp}. Such extreme conditions are anticipated to significantly influence the behavior of quark matter, potentially leading to novel phenomena, such as magnetic catalysis~\cite{Gusynin:1994re}, inverse magnetic catalysis~\cite{Preis:2010cq}, chiral magnetic effect, etc.~\cite{Kharzeev:2007jp,Fukushima:2008xe}.

Building on these experimental advances, it is essential to deepen the theoretical understanding of strong magnetic fields, particularly constant ones, given the transient nature of fields generated in colliders. Research in this area includes the Nambu-Jona-Lasinio (NJL) model~\cite{Boomsma:2009yk,Gatto:2012sp,Ferreira:2013tba,Ferreira:2014kpa,Chu:2014pja,Yu:2014xoa,Liu:2018zag,Xu:2020yag}, Holographic QCD~\cite{Bergman:2008sg,Johnson:2008vna,Filev:2009xp,Preis:2010cq,Preis:2012fh,Ballon-Bayona:2013cta,Mamo:2015dea}, Lattice QCD~\cite{Ejiri:2009ac,DElia:2010abb,Bali:2011qj,DElia:2011koc,Bali:2012zg,Bruckmann:2013oba,Bali:2013esa,Bali:2014kia,Endrodi:2015oba}, Functional Renormalization Group Method~\cite{Skokov:2011ib,Andersen:2012bq,Andersen:2013swa,Kamikado:2013pya,Kamikado:2014bua,Braun:2014fua}, and Dyson-Schwinger equations in both QED~\cite{Gusynin:1999pq,Leung:2005xz,Leung:2005yq,Rojas:2008sg,Ferrer:2008dy,Ferrer:2009nq} and QCD~\cite{Kojo:2012js,Mueller:2014tea,Mueller:2015fka}, among others. Theoretically, unlike electric fields, constant magnetic fields do not perform work on charged particles, enabling the analysis of equilibrated systems in their presence. This unique characteristic facilitates the study of matter dynamics and related phenomena within magnetic fields without the complexities of energy transferring by the field itself. From quantum mechanics, when an electron interacts with an external magnetic field, its energy levels in the plane perpendicular to the field become quantized into discrete Landau levels. However, along the direction parallel to the magnetic field, the electron's motion remains unconstrained, mimicking free-electron behavior. This dichotomy introduces a fundamental anisotropy in the response of the electron to the magnetic field~\cite{landau2013electrodynamics,landau2013quantum}.

Now considering QCD, quarks in an external magnetic field are expected to exhibit effects similar to those of electrons. However, in the infrared region, quarks are significantly influenced by gluon dressing, leading to an effective mass that exceeds the bare current quark mass, approaching a scale comparable to \(\Lambda_{\text{QCD}}\). This dressing effect causes quarks to behave as quasiparticles, characterized by an effective mass often referred to as the constituent quark mass~\cite{Roberts:1994dr}. Thus, when analyzing quarks in an external magnetic field, it is crucial to consider both the magnetic field's influence and the role of gluon dressing, particularly in the infrared domain. Extensive research has either overlooked quark-gluon interactions by treating quarks as ``free" particles~\cite{Ayala:2015bgv} or replaced the current quark mass with a momentum-independent constituent mass, a common approach in the NJL model~\cite{Fukushima:2012kc,Xu:2020yag}. Although these approaches provide a foundation for more complex analyses involving variables such as temperature and chemical potential~\cite{Cao:2015xja,Ayala:2016bbi,Sheng:2022ssp}, it is essential to prioritize accurate momentum-dependent gluon dressing, as dictated by the strong interaction.

Several methods, including Landau-level representation, the Schwinger proper-time formalism, and the Ritus method, have been developed to study the propagator of ``free" quarks in an external magnetic field. These representations, while distinct, are inter-convertible, providing different perspectives on the quark propagator. The Ritus method is particularly advantageous because of the simplified form of the ``free" quark propagator on the Ritus basis.  Additionally, it provides a straightforward framework for deriving both Landau-level and Schwinger proper-time representations from it. Thus, we begin by applying the gluon dressing to the ``free" quark propagator within the Ritus basis, deriving its general form in both coordinate and momentum spaces. Following this, we numerically solve the associated gap equation to obtain results for the dressed quark propagator. This method aligns with established studies on the fermion propagator using the Dyson-Schwinger equation in both QED~\cite{Gusynin:1999pq,Leung:2005xz,Leung:2005yq,Rojas:2008sg,Ferrer:2008dy,Ferrer:2009nq} and QCD~\cite{Kojo:2012js,Mueller:2014tea,Mueller:2015fka}.

The paper is organized as follows. Section \ref{sec:dispersion} discusses the energy dispersion relation of quarks in an external magnetic field. Section \ref{sec:quarkpro} presents the general form of the ``free" quark propagator and its dressed form in the weak-field approximation. Section \ref{sec:gap} addresses the gap equation that governs the propagator. Section \ref{sec:results} provides the numerical results, and Section \ref{sec:summary} concludes with a summary.

\section{Energy dispersion relation}\label{sec:dispersion}

\begin{figure}[t]
\includegraphics[width=0.92\columnwidth]{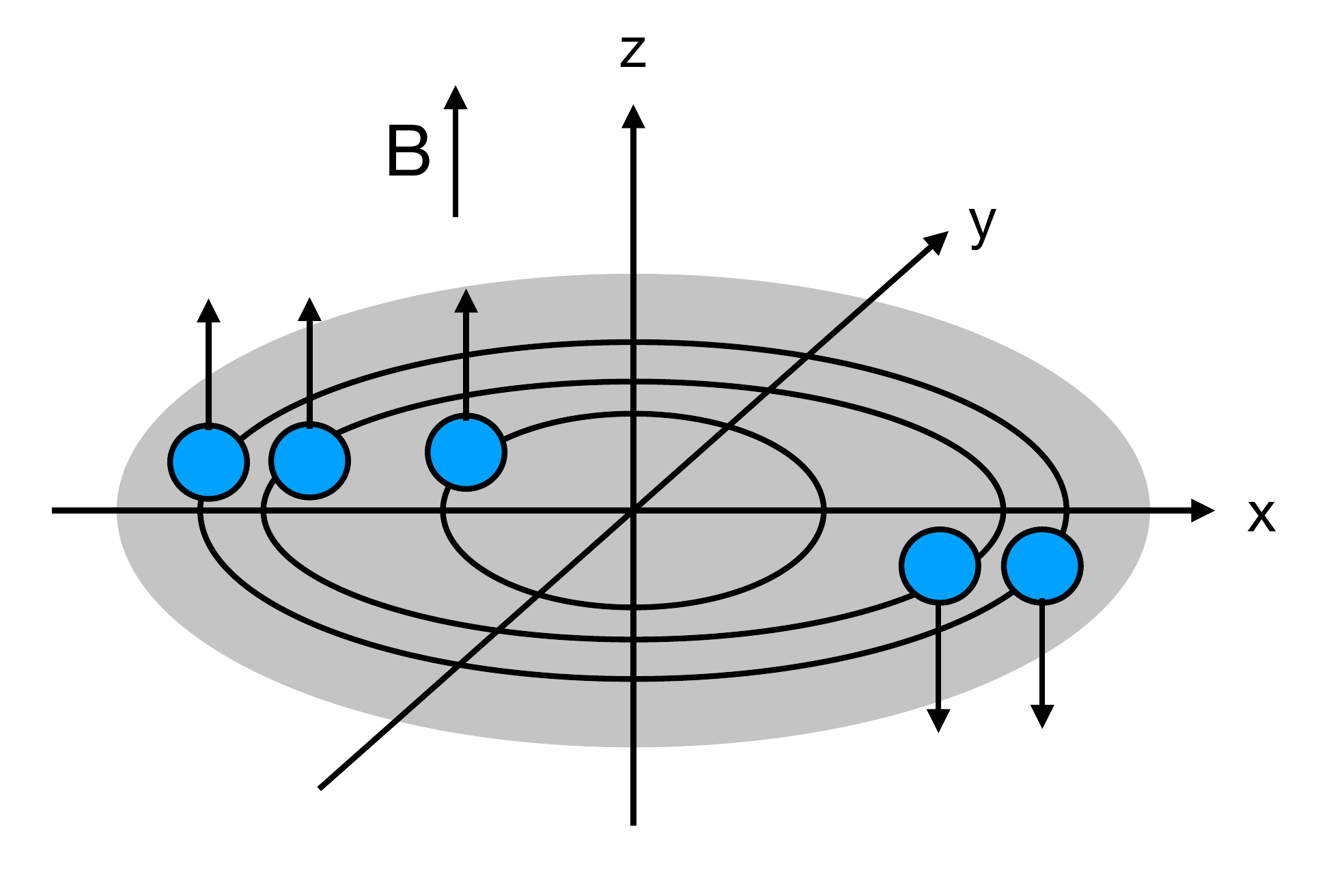}
\caption{Illustration of quarks in a magnetic field: Positively charged quarks are depicted as blue spheres, with arrows indicating their spin directions. Black ellipses represent discrete Landau energy levels.}
\label{fig:sketch}
\end{figure}

In the presence of a constant magnetic field $B$ oriented along the $z$-axis, the electromagnetic vector potential in the Landau gauge is given by $A^\mu_{\text{ext}}=(0,0,Bx,0)$. The energy dispersion relation for a quark in this field is expressed as:
\begin{align}
	\epsilon_n^2-p_z^2-(2n+1- 2s_z)|q_fB|-m^2=0\,,
\end{align}
where $q_f$ is the electric charge of the quark, $s_z=\pm1/2$ $(s_z=\mp 1/2)$ corresponds to the spin   parallel (anti-parallel) to the $z$-axis for positively (negatively) charged quarks. For positively charged quarks, when $n=0$, the spin aligns with the magnetic field ($s_z=1/2$), and the energy dispersion relation simplifies to $\epsilon_0^2=p_z^2+m^2$, representing the unique ground state or the lowest Landau level (LLL). For $n>0$, energy levels exhibit two-fold spin degeneracy due to the Zeeman effect. A schematic representation is provided in Fig.~\ref{fig:sketch}.

The dispersion relation reveals the distinct behaviors of quark motion parallel and transverse to a magnetic field. The longitudinal momentum $p_z$ remains continuous, allowing for unimpeded quark propagation parallel to the field. In contrast, the term $2n|q_fB|$ represents the quantized Landau levels that govern transverse motion, imposing constraints that restrict quarks to discrete energy states.\footnote{Note that there are two degrees of freedom in the transverse plane: the Landau level index $n$, which characterizes energy quantization, and the guiding center of the cyclotron orbits, linked to the conserved canonical momentum in the $y$-direction, $p_y$ (not explicitly present in the dispersion relation).} The additional term $-2s_z|q_fB|$ introduces spin-dependent energy splitting. Spin-up quarks, aligned with the magnetic field, occupy lower energy states, including the non-degenerate LLL. In contrast, spin-down quarks, anti-aligned with the field, occupy higher energy states, with higher Landau levels exhibiting two-fold degeneracy due to possible occupation by both spin orientations. 

The dispersion relation can be expressed equivalently as:
\begin{align}\label{eq:dispersion}
	\epsilon_n^2-p_z^2-2n|q_fB|-m^2=0\,,
\end{align}
where $n=0,1,2\dots$ corresponds to spin aligned with the $z$-axis and $n=1,2,3\dots$ corresponds to spin anti-aligned with the $z$-axis.

The two-fold spin degeneracy of the energy level suggests a new basis that incorporates the superposition of projection operators for the two spin states, known as the Ritus basis~\cite{Ritus:1972ky,Ritus:1978cj}. In this basis, the wave function of the Dirac equation is factorized into longitudinal and transverse components, allowing the spinor to satisfy the ``free" Dirac equation. The quark propagator, without quark-gluon interactions, is similarly simplified in the Ritus basis space, resembling the ``free" propagator:
\begin{align}\label{eq:ritusfreeinv}
	S_0^{-1}(p_n)&=\slashed{p}_n-m\,,
\end{align} 
where $p_n\equiv (\epsilon_n,0,\sqrt{2n|q_fB|},p_z)$. Rewriting this in terms of the longitudinal momentum $p_\parallel=(\epsilon_n,0,0,p_z)$ yields:
\begin{align}
	S_0^{-1}(p_\parallel,n)=\slashed{p}_\parallel-\sqrt{2n|q_fB|}\gamma_2-m\,.
\end{align}
Taking the inverse of this expression allows us to derive the energy dispersion relation from the denominator of the propagator:
\begin{align}
	p_\parallel^2-2n|q_fB|-m^2=0\,,
\end{align} 
which corresponds precisely to the dispersion relation in Eq.~\eqref{eq:dispersion}.

When quark-gluon interactions are included, the quark propagator is dressed by gluons, complicating the dynamics of quark motion in a magnetic field. The Dirac projection matrices
\begin{align}
	\Sigma^\pm=\frac{1}{2}\left(\mathds{1}\pm \Sigma^3\right)\,,\quad \Sigma^3=i\gamma^1\gamma^2\,,
\end{align}
serve as spin projection operators to isolate the spin-up and spin-down states. Using these matrices, the general expression for the quark propagator in a magnetic field, within the Ritus basis, is formulated as follows~\cite{Watson:2013ghq}:
\begin{align}\label{eq:ritusdressinv}
	S^{-1}(p_\parallel,n) ={}&\Sigma^+(\mathcal{A}\slashed{p}_\parallel-\mathcal{B})+\Sigma^-(\mathcal{C}\slashed{p}_\parallel-\mathcal{D})\notag\\
	&-\sqrt{2n|q_fB|}\gamma_2 \mathcal{E}\,,
\end{align}
with $\mathcal{A}$ to $\mathcal{E}$ are dressing functions dependent on $p_\parallel$ and $n$. Since quarks propagate unimpeded along the magnetic field, the projection operators $\Sigma^\pm$ are used to separate contributions from spin-up and spin-down states, with $\Sigma^+$ corresponding to spin-down and $\Sigma^-$ to spin-up. The longitudinal propagation of the quark is dressed similarly to the vacuum case. Due to Zeeman splitting, spin-up and spin-down quarks have distinct effective energies, requiring different dressing functions for their longitudinal propagation: $\mathcal{A} \slashed{p}_\parallel - B$ for spin-down quarks and $\mathcal{C} \slashed{p}_\parallel - D$ for spin-up quarks. Notably, in the LLL, only spin-up quarks contribute, simplifying the propagator to $\Sigma^-(\mathcal{C}\slashed{p}_\parallel-\mathcal{D}$). The transverse propagation of quarks, which involves a mixture of spin-up and spin-down states, is constrained to quantized Landau levels. When quark-gluon interactions are included, this transverse motion is further modified by the dressing function $\mathcal{E}$.

Alternatively, the inverse propagator in Eq.~\eqref{eq:ritusdressinv} can be expressed in terms of $\Sigma^3$ as follows:
\begin{align}
	S^{-1}(p_\parallel,n) 
 ={}&-\mathcal{S}+\mathcal{V}_\parallel\slashed{p}_\parallel-\sqrt{2n|q_fB|}\gamma_2 \mathcal{V}_\perp\notag\\
&+|q_fB|\mathcal{A}_v\Sigma^3\slashed{p}_\parallel-2|q_fB|\mathcal{T}\Sigma^3\,,
\end{align}
with $\mathcal{S}$, $\mathcal{V}_\parallel$, $\mathcal{V}_\perp$, $\mathcal{A}_v$ and $\mathcal{T}$ are re-combinations of the dressing functions $\mathcal{A}$ to $\mathcal{E}$:
\begin{align*}
    \mathcal{S}&=\frac{\mathcal{B}+\mathcal{D}}{2}\,,&\mathcal{V}_\parallel&=\frac{\mathcal{A}+\mathcal{C}}{2}\,,& \mathcal{V}_\perp=\mathcal{E}\,, \notag\\
    |q_fB|\mathcal{A}_v&=\frac{\mathcal{A}-\mathcal{C}}{2}\,,&2|q_fB|\mathcal{T}&=\frac{\mathcal{B}-\mathcal{D}}{2}\,.&{\addtocounter{equation}{1}\tag{\theequation}}
\end{align*}
This dressed propagator illustrates that the dressing functions in the parallel and perpendicular directions to the magnetic field, $\mathcal{V}_\parallel$ and $\mathcal{V}_\perp$, are fundamentally different, reflecting the anisotropy of the system under an external magnetic field. In addition, two new terms arise: an axial-vector term $|q_fB|\mathcal{A}_v\Sigma^3\slashed{p}_\parallel$ and a tensor term $2|q_fB|\mathcal{T}\Sigma^3$. We will further explore their physical implications and interpretations.

Taking the inverse of the inverse propagator in Eq.~\eqref{eq:ritusdressinv} yields a dressed version of the dispersion relation:
\begin{align}\label{eq:dispersiondress}
    p_\parallel^2=&\frac{\Omega_1   +\Omega_2-\Omega_3\pm2|q_fB|\left(\mathcal{A}_v\mathcal{S}-2\mathcal{V}_\parallel\mathcal{T}\right)\sqrt{\Omega_4}}{\left(\mathcal{V}_\parallel^2-|q_fB|^2\mathcal{A}_v^2\right)^2}\,,
\end{align}
where $\Omega_{1-4}$ are defined in \cref{appendix:inverseRitus}. Apparently, the non-zero dressing functions associated with the magnetic-field-induced axial-vector and tensor terms lead to energy splitting between spin states within the same Landau level, directly corresponding to the Zeeman effect. 
In the specific limit where where 
\begin{align}  \mathcal{V}_\parallel=\mathcal{V}_\perp=1\,,\quad \mathcal{A}_v=0\,,\quad \mathcal{S}=M\,,
\end{align}
as used in Ref.~\cite{Ferrer:2008dy}, we find: 
\begin{align}
\Omega_1&=0\,,\notag\\
\Omega_2&=M^2+4|q_fB|^2\mathcal{T}^2+2n|q_fB|\,,\notag\\
\Omega_3&=0\,,\notag\\
\Omega_4&=M^2+2n|q_fB|\,.
\end{align}
This leads to:
\begin{align}
p_\parallel^2
&=\left(\sqrt{M^2+2n|q_fB|}\pm 2|q_fB|\mathcal{T}\right)^2\,.
\end{align}
Clearly, the magnetic-field-induced tensor term disrupts the energy degeneracy between spin states within the same Landau level.

\section{Quark propagator}\label{sec:quarkpro}

The preceding discussion clearly demonstrates that the system's anisotropy is intrinsically linked to the quark propagator. Therefore, in the following sections we will systematically study the quark propagator under the influence of an external magnetic field.

\subsection{Schwinger Phase}

The  propagator and its inverse in coordinate space can be factorized as follows:
\begin{subequations}\label{eq:factorization1}
\begin{align}
 {S}(x,y) &= e^{i\Phi(x,y)} \bar{S}(x-y)\,,\\
 {S}^{-1}(x,y) &= e^{i\Phi(x,y)} \tilde{S}(x-y)\,,
\end{align}
\end{subequations}
where $\Phi(x,y)=\frac{q_fB}{2}(x_2-y_2)(x_1+y_1)$ is the Schwinger phase. The factorization presented in Eq.~\eqref{eq:factorization1} is universal  under certain approximations~\cite{Schwinger:1951nm,Gusynin:1999pq}, and here we will adopt this form. The Schwinger phase explicitly depends on the gauge field and breaks the translational invariance in the transverse plane. 
The terms $\bar{S}(x-y)$ and $\tilde{S}(x-y)$ represent the translational invariant parts of the propagator.
Referring to Eq.~\eqref{eq:factorization1}, we note that
\begin{align}
\tilde{S}(x-y)\neq\frac{1}{\bar{S}(x-y)}\,.
\end{align}
The factorization allows one to focus on the Fourier transformation of the translation-invariant part, converting both the translation-invariant parts of the propagator and its inverse from coordinate space to momentum space, yielding $\bar{S}(p_\parallel, p_\perp)$ and $\tilde{S}(p_\parallel, p_\perp)$:
\begin{subequations}\label{eq:factorization2}
\begin{align}
  \bar{S}(x-y) &= \int \frac{d^4 p}{(2\pi)^4} e^{-i p \cdot (x-y)} \bar{S}(p_\parallel, p_\perp)\,,\\
  \tilde{S}(x-y) &= \int \frac{d^4 p}{(2\pi)^4} e^{-i p \cdot (x-y)} \tilde{S}(p_\parallel, p_\perp)\,.
\end{align}
\end{subequations}
Consequently, in momentum space, we find:
\begin{align}
\tilde{S}(p_\parallel, p_\perp)\neq\frac{1}{\bar{S}_1(p_\parallel, p_\perp)}\,.
\end{align}
Typically, we denote $\bar{S}(p_\parallel, p_\perp)=S(p_\parallel, p_\perp)$ and $\tilde{S}(p_\parallel, p_\perp)=S^{-1}(p_\parallel, p_\perp)$, leading to:
\begin{align}\label{eq:counterintuitive}
  S^{-1}(p_\parallel, p_\perp)\neq \frac{1}{S(p_\parallel, p_\perp)}\,.
\end{align}
At first glance, this relation may seem counterintuitive; however, its apparent peculiarity arises from the factorization of the Schwinger phase, which will show great simplicity in solving the quark gap equation.

\subsection{Free Propagator in a Magnetic Field}
 
The propagator of a free quark in an external magnetic field can be represented in three equivalent forms: Landau-level, Schwinger proper-time, and Ritus-basis representations~\cite{Hattori:2023egw}. In the Landau-level representation (restricted to $h\equiv q_fB\geq0$), the propagator and its inverse in momentum space are given by:
\begin{align}
    S_0^{-1}(p_\parallel, p_\perp) = {}&  e^{-p_\perp^2/h} \sum_{n=0}^\infty 2(-1)^n \notag \\
    & \times\Biggl\{ \left[ -\Sigma^+ L_{n-1}^{0} \left( \alpha_\perp\right) + \Sigma^- L_n^{0} \left( \alpha_\perp \right) \right] \notag \\
    & \times \left( \slashed{p}_\parallel - m \right) + 2 L_{n-1}^{1} \left(\alpha_\perp\right) \slashed{p}_\perp \Biggl\}\,,\notag\\
    S_0(p_\parallel, p_\perp) = {}&  e^{-p_\perp^2/h} \sum_{n=0}^\infty \frac{2(-1)^n}{p_\parallel^2-2nh-m^2+i0_+} \notag \\
    & \times\Biggl\{ \left[ -\Sigma^+ L_{n-1}^{0} \left( \alpha_\perp \right) + \Sigma^- L_n^{0} \left( \alpha_\perp \right) \right] \notag \\
    & \times \left( \slashed{p}_\parallel + m \right) + 2 L_{n-1}^{1} \left( \alpha_\perp \right) \slashed{p}_\perp \Biggl\}\,.
\end{align}
Here, $\alpha_\perp=\frac{2}{h} p_\perp^2$, and $L_n^\alpha$ are the Laguerre polynomials with $p_\perp=(0,p_x,p_y,0)$ representing the transverse momentum. The terms involving $\Sigma^+(\slashed{p}_\parallel\pm m)$ and $\Sigma^-(\slashed{p}_\parallel\pm m)$ correspond to the longitudinal propagation of the spin-down and spin-up quarks, respectively. The term $\slashed{p}_\perp$ accounts for transverse propagation, incorporating a mixing of both spin-up and spin-down states~\cite{Hattori:2023egw}.

The summation over Landau levels can be performed analytically, leading to the Schwinger proper-time representation:
\begin{align}\label{eq:Schwingerfree}
    &S_0^{-1}(p_\parallel, p_\perp) = \slashed{p} - m\,,\notag\\
    &S_0(p_\parallel, p_\perp)\notag\\
&=-i\int_0^\infty ds\, \text{exp}\left[is\left({p}_\parallel^2-m^2+i0_+\right)-i\frac{p_\perp^2}{h}\tan(sh)\right]\notag\\
	&\times\left[\slashed{p}+m+ \gamma_1\gamma_2(\slashed{{p}}_\parallel+m)\tan(sh)
	-\slashed{p}_\perp\tan^2(sh)\right]\,.
\end{align}
The inverse propagator remarkably resembles its form in the absence of a magnetic field, with the only modification being the introduction of the Schwinger phase. However, deriving the propagator itself requires integration in Schwinger's proper time $s$. As indicated in Eq.~\eqref{eq:Schwingerfree}, the relationship between the propagator and its inverse in momentum space is nontrivial:
\begin{align}
  S^{-1}_0(p_\parallel, p_\perp)\neq \frac{1}{S_0(p_\parallel, p_\perp)}\,,      
\end{align}
as noted in Eq.~\eqref{eq:counterintuitive}.
This distinction is critical; in the absence of a magnetic field, it is often assumed that this equality holds without question. However, this assumption does not apply when a magnetic field is present. As we proceed to solve the gap equation, it is important to recognize that the propagator appears on the right-hand side of the equation. Given the form of the inverse propagator, we cannot simply take its inverse to obtain the propagator. Instead, we must derive the propagator using the Ritus basis expansion, as outlined below.

The third representation employs the Ritus eigenfunction method, which replaces the traditional four-dimensional momentum space with two parallel momentum components $p_\parallel$ and the Landau level index $n$. Using the Ritus basis, the propagator and its inverse in coordinate space are expressed as~\cite{Ritus:1972ky,Ritus:1978cj}:
\begin{align}\label{eq:ritusbasisexpansion}
    S_0^{\kappa}(x, y) = \sum_{n=0}^\infty \int \frac{d^3 \tilde{p}}{(2\pi)^3} E(x; \tilde{p}, n) S_0^{\kappa}(p_n) \bar{E}(y; \tilde{p}, n)\,,
\end{align}
where $\tilde{p}=(p_0,0,p_2,p_3)$, $\kappa=-1,1$. $S_0^{-1}(p_n)$ represents the inverse propagator in the Ritus-basis space as defined in Eq.~\eqref{eq:ritusfreeinv}. $S_0(p_n)$ represents the propagator in the Ritus-basis space. The Ritus basis provides a significant advantage: the relationship between the propagator and its inverse in Ritus-basis space is straightforward, 
\begin{align}\label{eq:ritusdresspro}
	S_0^{-1}(p_n)=\frac{1}{S_0(p_n)}\,,
\end{align}
allowing us to express the propagator as
\begin{align}\label{eq:ritusfree}
	S_0(p_n)
	=\frac{\slashed{p}_n+m}{p_n^2-m^2+i0_+}\,,
\end{align}
where $p_n^2=p_\parallel^2-2nh$. This method employs orthonormal and complete Ritus matrices, $E$ and $\bar{E}$, instead of the traditional Fourier exponential form $e^{-ip \cdot x}$and $e^{ip \cdot x}$ to expand the quark propagator:
\begin{align}
    E(x; \tilde{p}, n) &= h^{1/4} e^{-i \tilde{p} \cdot x} \left[ \psi_{n-1}(\varepsilon) \Sigma^+ + \psi_n(\varepsilon) \Sigma^- \right], \notag \\
    \bar{E}(y; \tilde{p}, n) &= h^{1/4} e^{i \tilde{p} \cdot y} \left[ \psi_{n-1}(\tau) \Sigma^+ + \psi_n(\tau) \Sigma^- \right],
\end{align}
where $\psi_n$ denotes the Hermite function indexed by $n$, with arguments
    $\varepsilon = \sqrt{h} x_1 + \frac{p_2}{\sqrt{h}}$ and $\tau = \sqrt{h} y_1 + \frac{p_2}{\sqrt{h}}$. Hermite functions are versatile and have recently shown a significant correspondence between free-field and interacting states~\cite{Gao:2022ylj}. In the LLL, where only $n=0$ mode is relevant, the characteristic Ritus matrix structures reduce to $\psi_0\Sigma^-$, given that $\psi_{-1}=0$. The Ritus matrices are orthonormal and complete:
\begin{align}
    \int d^4 x \, \bar{E}(x; \tilde{p}, n) E(x; \tilde{q}, m) &= \delta_{nm} (2\pi)^3 \delta(\tilde{p} - \tilde{q}) \tilde{\mathds{1}}_n, \notag \\
    \sum_{n=0}^\infty \int \frac{d^3 \tilde{p}}{(2\pi)^3} E(x; \tilde{p}, n) \bar{E}(y; \tilde{p}, n) &= \delta(x - y),
\end{align}
where $\tilde{\mathds{1}}_n$ is defined as:
\begin{align}
    \tilde{\mathds{1}}_n =
    \begin{cases}
        \Sigma^-, & n = 0 \\
        \mathds{1}, & n > 0
    \end{cases},
\end{align}
and it is evident that only $\Sigma^-$ contributes to the LLL.

All three representations are equivalent and can easily be derived from one another. However, it is beneficial to choose the representation that maintains a straightforward structure, facilitating the analysis of the propagator after incorporating quark-gluon interactions. The Ritus-basis representation is particularly advantageous in this regard, as both the inverse propagator and the propagator in Ritus-basis space exhibit simple forms that resemble the free propagator,with four-momentum replaced by $p_n$, as shown in Eqs.~\eqref{eq:ritusfreeinv} and ~\eqref{eq:ritusfree}. Therefore, we will employ the Ritus-basis representation to study the quark propagator dressed by gluons in an external magnetic field.

\subsection{Dressed Inverse Propagator in a Magnetic Field}

When quark-gluon interactions are included, the inverse propagator in the Ritus basis space is dressed, transitioning from Eq.~\eqref{eq:ritusfreeinv} to Eq.~\eqref{eq:ritusdressinv}. Substituting Eq.~\eqref{eq:ritusdressinv} into the Ritus-basis representation in Eq.~\eqref{eq:ritusbasisexpansion} yields the factorized form of the propagator in Eq.~\eqref{eq:factorization1}, with the inverse propagator in momentum space given by:
\begin{align}\label{eq:inverse}
	S^{-1}(p_\parallel,p_\perp)={}&e^{-p_\perp^2/h}
	\sum_{n=0}^\infty 2(-1)^{n}\notag\\
	&\times\biggl[-\Sigma^+(\mathcal{A}\slashed{p}_\parallel - \mathcal{B})L_{n-1}^{0}\left(\alpha_\perp\right)\notag\\
	&+\Sigma^-(\mathcal{C}\slashed{p}_\parallel - \mathcal{D})L_n^{0}\left(\alpha_\perp\right)\notag\\
	&+2\slashed{p}_\perp \mathcal{E} L_{n-1}^{1}\left(\alpha_\perp\right)\biggl]\,.
\end{align}
If we make the following identifications for the summations over Landau levels involving the dressing functions~\cite{Watson:2013ghq}:
\begin{align}
	e^{-p_\perp^2/h}\sum_{n=0}^\infty 2(-1)^{n-1}\mathcal{A}({p}_\parallel,n)L_{n-1}^{0}\left(\alpha_\perp\right)&=\hat{A}({p}_\parallel^2,p_\perp^2)\,,\notag\\
	e^{-p_\perp^2/h}\sum_{n=0}^\infty 2(-1)^{n-1}\mathcal{B}({p}_\parallel,n)L_{n-1}^{0}\left(\alpha_\perp\right)&=\hat{B}({p}_\parallel^2,p_\perp^2)\,,\notag\\
	e^{-p_\perp^2/h}\sum_{n=0}^\infty 2(-1)^{n}\mathcal{C}({p}_\parallel,n)L_{n}^{0}\left(\alpha_\perp\right)&=\hat{C}({p}_\parallel^2,p_\perp^2)\,,\notag\\
	e^{-p_\perp^2/h}\sum_{n=0}^\infty 2(-1)^{n}\mathcal{D}({p}_\parallel,n)L_{n}^{0}\left(\alpha_\perp\right)&=\hat{D}({p}_\parallel^2,p_\perp^2)\,,\notag\\
	e^{-p_\perp^2/h}\sum_{n=0}^\infty 4(-1)^{n-1}\mathcal{E}({p}_\parallel,n)L_{n-1}^{1}\left(\alpha_\perp\right)&=\hat{E}({p}_\parallel^2,p_\perp^2)\,,	
\end{align}
the inverse propagator transforms into:
\begin{align}\label{eq:inversestructures}
	&S^{-1}(p_\parallel,p_\perp)=\Sigma^+( \hat{A}\slashed{p}_\parallel- \hat{B})+\Sigma^-( \hat{C} \slashed{p}_\parallel- \hat{D})
	-\slashed{p}_\perp \hat{E}\,.
\end{align}
The functions $\hat{A}$ to $\hat{E}$ depend on ${p}_\parallel^2$ and $p_\perp^2$, encapsulating the implicit dependence on $h$. The terms involving $\Sigma^+( \hat{A}\slashed{p}_\parallel- \hat{B})$ and $\Sigma^-(\hat{C} \slashed{p}_\parallel- \hat{D})$ correspond to the longitudinal propagation of spin-down and spin-up quarks, respectively, while the term $\slashed{p}_\perp \hat{E}$ represents the transverse propagation of quarks, incorporating a mixing of both spin states. Alternatively, using $\Sigma^3$ instead of $\Sigma^\pm$ gives us:
\begin{align}\label{eq:16diracbasis}
	S^{-1}(p_\parallel,p_\perp)
	&=-\hat{\mathbb{S}} + \hat{\mathbb{V}}_\parallel \slashed{p}_\parallel - \hat{\mathbb{V}}_\perp \slashed{p}_\perp +h\hat{\mathbb{A}}  \Sigma^3\slashed{p}_\parallel -2h \hat{\mathbb{T}}\Sigma^3\,,
\end{align}
where $\hat{\mathbb{S}}$, $\hat{\mathbb{V}}_\parallel$, $\hat{\mathbb{V}}_\perp$, $\hat{\mathbb{A}}$ and $\hat{\mathbb{T}}$ are re-combinations of $\hat{A}$ to $\hat{E}$:
\begin{align*}
	\hat{\mathbb{S}}&=\frac{\hat{B}+\hat{D}}{2}\,,
	& \hat{\mathbb{V}}_\parallel &=\frac{\hat{A}+\hat{C}}{2}\,,
	& \hat{\mathbb{V}}_\perp &=\hat{E}\,,\notag\\
	h\hat{\mathbb{A}}&=\frac{\hat{A}-\hat{C}}{2}\,,& 2h\hat{\mathbb{T}}&=\frac{\hat{B}-\hat{D}}{2}\,.{\addtocounter{equation}{1}\tag{\theequation}}\label{eq:AEtoST}
\end{align*}
In \cref{appendix:structure}, we demonstrate that this formulation comprehensively captures the Dirac structures relevant to the quark propagator in an external magnetic field. In particular, the vector term is decomposed into longitudinal ($\hat{\mathbb{V}}_\parallel \slashed{p}_\parallel$) and transverse ($\hat{\mathbb{V}}_\perp \slashed{p}_\perp$) components relative to the magnetic field direction. Additionally, as shown in Eq.~\eqref{eq:AEtoST}, $\hat{\mathbb{S}}$, and $\hat{\mathbb{V}}_\parallel \slashed{p}_\parallel$ correspond to the effective mass and longitudinal momentum for the averaged spin states.

Furthermore, the axial-vector term ($h\hat{\mathbb{A}}  \Sigma^3\slashed{p}_\parallel$) and the tensor term ($2h \hat{\mathbb{T}}\Sigma^3$) emerge as novel contributions in the presence of a magnetic field, vanishing when the magnetic field is absent. Both terms are coupled with $\Sigma^3$, reflecting their role in the description of fine structures arising from interactions between quark spin and the external magnetic field, showing in more detail the subtle spin-dependent behavior in the quark propagator. Specifically, as shown in Eq.~\eqref{eq:AEtoST}, the axial-vector term captures the asymmetry in longitudinal momentum between spin-up and spin-down states, while the tensor term accounts for asymmetry in masses between these states, which can be interpreted as a non-perturbative Zeeman effect. Additionally, the tensor term introduces further chiral symmetry breaking, complementing that from the scalar term $\hat{\mathbb{S}}$. 

Interestingly, two distinct effective masses can be defined for the quark in a magnetic field: the longitudinal mass $\hat{\mathbb{M}}^{\text{eff}}_\parallel$ and the transverse mass $\hat{\mathbb{M}}^{\text{eff}}_\perp$, given by:
\begin{align}\label{eq:Mhnonzero}
\hat{\mathbb{M}}^{\text{eff}}_\parallel=\frac{\hat{\mathbb{S}}}{\hat{\mathbb{V}}_\parallel}\,,\quad\text{and}\quad \hat{\mathbb{M}}^{\text{eff}}_\perp=\frac{\hat{\mathbb{S}}}{\hat{\mathbb{V}}_\perp}\,.
\end{align}
These effective masses characterize the quark's mass response to an external magnetic field. In the absence of a magnetic field ($h=0$), the dressing functions satisfy:
\begin{align}\label{eq:AEh0}
	\hat{{A}}=\hat{{C}}=\hat{{E}}\,,\quad\quad \hat{{B}}&=\hat{{D}}\,,
\end{align}
or equivalently,
\begin{align}\label{eq:STh0}
	\hat{\mathbb{V}}_\parallel &=\hat{\mathbb{V}}_\perp\,.
\end{align}
Under these conditions, the propagator simplifies to the standard form $S^{-1}(p)=\hat{{A}}\slashed{p}-\hat{{B}}$ and the two effective masses become identical:
\begin{align}\label{eq:Mh0}
\hat{\mathbb{M}}^{\text{eff}}_\parallel=\hat{\mathbb{M}}^{\text{eff}}_\perp\,.
\end{align}
This indicates isotropic quark propagation in all directions. We will further explore how the effective masses behave in different directions when an external magnetic field is applied.

\subsection{Dressed Propagator in a Weak Magnetic Field}




Starting from the definition of the propagator and its inverse form as presented in Eq.~\eqref{eq:factorization1} and using the inverse propagator defining in Eq.~\eqref{eq:inverse}, we derive the expression for the propagator in momentum space as follows:
\begin{align}
	S(p_\parallel,p_\perp)={}&e^{-p_\perp^2/h}
	\sum_{n=0}^\infty 2(-1)^{n}\notag\\
	&\times\biggl[-\Sigma^+(\mathcal{W_A}\slashed{p}_\parallel + \mathcal{W_B})L_{n-1}^{0}\left(\alpha_\perp\right)\notag\\
	&+\Sigma^-(\mathcal{W_C}\slashed{p}_\parallel + \mathcal{W_D})L_n^{0}\left(\alpha_\perp\right)\notag\\
	&+2\slashed{p}_\perp \mathcal{W_E} L_{n-1}^{1}\left(\alpha_\perp\right)\biggl]\,.
\end{align}

The scalar functions are defined as follows:
\begin{align*}
	\mathcal{W_A}=&\frac{\Delta_1\mathcal{C}-\Delta_2 \mathcal{D}}{\Delta}\,,&\mathcal{W_B}=&\frac{\Delta_1\mathcal{D}-p_\parallel^2\mathcal{C}\Delta_2}{\Delta}\,,\\	\mathcal{W_C}=&\frac{\Delta_1\mathcal{A}+\Delta_2 \mathcal{B}}{\Delta}\,,&\mathcal{W_D}=&\frac{\Delta_1\mathcal{B}+p_\parallel^2\mathcal{A}\Delta_2}{\Delta}\,,\\
	\mathcal{W_E}=&\frac{\Delta_1\mathcal{E}}{\Delta}\,.{\addtocounter{equation}{1}\tag{\theequation}}
\end{align*}
The denominator $\Delta$ is defined as $\Delta =\Delta_1^2-p_\parallel^2\Delta_2^2$, where $\Delta_1$ and $\Delta_2$ are detailed in Eq.~\eqref{eq:denominator}.
Due to the complex $n$-dependence in the denominator, summing over Landau levels is highly nontrivial. As suggested in Ref.~\cite{Watson:2013ghq}, this summation can be performed in the weak field limit as $h\rightarrow 0$, where $\Delta_2$ in the expressions for $\mathcal{W_A}$ to $\mathcal{W_E}$ vanishes:
\begin{align}
	\Delta_2\overset{h\rightarrow0}{\longrightarrow}0\,.
\end{align}
Expanding the denominator $\Delta$ in terms of $\Delta_2$ gives:
\begin{align}
	\frac{1}{\Delta}=\frac{1}{\Delta_1^2}+\mathcal{O}(\Delta_2^2)\,.
\end{align}
Keeping only the first-order term in $\Delta_2$, the scalar functions: $\mathcal{W_A}$ to $\mathcal{W_E}$ become
\begin{align*}
	\mathcal{W_A}=&\frac{\mathcal{C}}{\Delta_1}-\frac{\Delta_2 \mathcal{D}}{\Delta_1^2}\,, & \mathcal{W_B}=\frac{\mathcal{D}}{\Delta_1}-\frac{{p}_\parallel^2\Delta_2 \mathcal{C}}{\Delta_1^2}\,,\\
	\mathcal{W_C}=&\frac{\mathcal{A}}{\Delta_1}+\frac{\Delta_2 \mathcal{B}}{\Delta_1^2}\,,&
	\mathcal{W_D}=\frac{\mathcal{B}}{\Delta_1}+\frac{{p}_\parallel^2\Delta_2 \mathcal{A}}{\Delta_1^2}\,,\\
	\mathcal{W_E}=&\frac{\mathcal{E}}{\Delta_1}\,.{\addtocounter{equation}{1}\tag{\theequation}}
\end{align*}
Notably, in the weak field limit, the propagator's denominator in the Ritus basis transitions from $\Delta$ to $\Delta_1$.

The summation over Landau levels can be performed, resulting in an expression involving an integral over the Schwinger proper time $s$, which includes the function $\tan(sh)$.
Expanding to the first order in $h$, we have:
\begin{align}
	\tan(sh)=sh+\mathcal{O}(h^3)\,.
\end{align}
The propagator then becomes:
\begin{widetext}
\begin{align}\label{eq:prostructures}
S(p_\parallel,p_\perp)
={}&\frac{\Sigma^+\left(\slashed{p}_\parallel \hat{C}+\hat{D}\right)+\Sigma^-\left(\slashed{p}_\parallel \hat{A} +\hat{B}\right)-\slashed{p}_\perp \hat{E} }{{p}_\parallel^2{\hat{A}}{\hat{C}} - p_\perp^2 \hat{E}^2-{\hat{B}}{\hat{D}}+i0_+} 
+\frac{h\hat{E}^2\left[\Sigma^+\left(\slashed{p}_\parallel  \hat{C}+ \hat{D}\right)-\Sigma^-\left(\slashed{p}_\parallel \hat{A} +\hat{B}\right)\right]}{({p}_\parallel^2{\hat{A}}{\hat{C}} - p_\perp^2 \hat{E}^2-{\hat{B}}{\hat{D}}+i0_+)^2}\notag\\
&+\frac{(\hat{A}\hat{D}-\hat{B}\hat{C})\left[-\Sigma^+\left(\slashed{{p}}_\parallel  \hat{D}+ {p}_\parallel^2 \hat{C}\right)
	+\Sigma^-\left(\slashed{p}_\parallel  \hat{B} +{p}_\parallel^2 \hat{A} \right)\right]}{({p}_\parallel^2{\hat{A}}{\hat{C}} - p_\perp^2 \hat{E}^2-{\hat{B}}{\hat{D}}+i0_+)^2}\,.
\end{align}
\end{widetext}
In particular, the propagator in momentum space is expressed in terms of the scalar functions $\hat{A}$ to $\hat{E}$ derived from the inverse propagator. Note that the general Dirac structure of the propagator aligns precisely with the previously derived form albeit with modifications in the scalar functions, which proves that it constitutes the complete basis for the propagator.

\section{The gap equation}\label{sec:gap}

The preceding discussion on the structure of the free and dressed-quark propagator is general and applicable to any analysis involving a constant magnetic field. 
To obtain the dressing functions from $\hat{A}$ to $\hat{E}$, we must resort to the gap equation for the quark propagator, which in configuration space is given by~\cite{Roberts:1994dr}:
\begin{align}\label{eq:gapeqcoordinate}
	S^{-1}(x,y)=S^{-1}_0(x,y)+g^2C_F\gamma^\mu S(x,y)\gamma^\nu D_{\nu\mu}(y,x)\,.
\end{align}
Here, $g$ denotes the QCD coupling constant and $C_F=4/3$ is the color factor.   We employ the rainbow truncation. $S^{-1}_0(x,y)$ is the free inverse propagator in the presence of a magnetic field, without gluon dressing.

The gluon propagator in the Landau gauge, denoted as $D_{\nu\mu}(y,x)$, can be expressed as:
\begin{align}
	D^{(0)}_{\nu\mu}(y,x)=\int\frac{d^4q}{(2\pi)^4}e^{-iq(y-x)}D^{(0)}_{\nu\mu}(q)\,,
\end{align}
with the momentum space representation given by:
\begin{align}
	D^{(0)}_{\nu\mu}(q)=\frac{1}{q^2}t_{\nu\mu}\,,
\end{align}
where $t_{\nu\mu}$ is the transverse momentum projector:
\begin{align}
	t_{\nu\mu}=g_{\nu\mu}-\frac{q_\nu q_\mu}{q^2}\,.
\end{align}
We assume that the gluon propagator is unaffected by the magnetic field, a hypothesis that can be refined in future studies by considering the magnetic field's impact on the gluon propagator. For this analysis, we used the vacuum-gluon propagator. In vacuum, the dressed gluon propagator can be expressed as:
\begin{align}
	g^2D_{\nu\mu}(q)=D^{(0)}_{\nu\mu}(q)q^2\mathcal{G}(q^2)\,,
\end{align}
where $\mathcal{G}(q^2)$ is the gluon dressing function that incorporates the QCD coupling. This function typically comprises two components:
\begin{align}
	\mathcal{G}(q^2)=\mathcal{G}_{\text{IR}}(q^2)+\frac{4\pi}{q^2}\alpha_{\text{pQCD}}(q^2)\,.
\end{align}
The ultraviolet part is given by:
\begin{align}
	\alpha_{\text{pQCD}}(q^2)=\frac{2\pi\gamma_m (1-e^{-q^2/[4m_t^2]})}{\ln[\tau+(1+q^2/\Lambda^2_{\text{QCD}})^2]}\,,
\end{align}
where $\gamma_m=12/(33-2N_f)$, $N_f=5$; $\Lambda_{\text{QCD}}=0.36$ GeV; $\tau=e^2-1$, $m_t=0.5$ GeV. For the infrared part, we use a model from Ref.~\cite{Qin:2011dd}:
\begin{align}
	\mathcal{G}_{\text{IR}}(q^2)=\frac{8\pi^2}{\omega^4}De^{-q^2/\omega^2}\,.
\end{align}
Empirical evidence suggests that the observable properties of light pseudoscalar and vector mesons composed of light quarks ($u$, $d$, $s$) remain largely stable by variations in $\omega$ within the range $\omega \in [0.4, 0.6]$ GeV, provided that the relation 
\begin{align}
	D\omega = (0.80\,\text{GeV})^3\,
\end{align}
is maintained. In this study, we explore potential variations using the values of $\omega = 0.4$, $0.5$, and $0.6$ GeV.

By substituting the factorized forms of the quark propagator and its inverse in Eq.~\eqref{eq:factorization1}, along with their Fourier transformation in Eq.~\eqref{eq:factorization2} into the gap equation in Eq.~\eqref{eq:gapeqcoordinate}, we obtain ($q=k-p$):
\begin{align}\label{eq:gapeqmom}
	S^{-1}(p) =& Z_2 (\slashed{p} - Z_m m) \notag\\ 
	& + Z_2^2 g^2 C_F \int \frac{d^4k}{(2\pi)^4} \gamma^\mu S(k) \gamma^\nu D_{\nu\mu}(q)\,.
\end{align}
Here, we employ a mass-independent momentum subtraction renormalization scheme, which, using the scalar Ward-Takahashi identity, determines the renormalization constants in the chiral limit, with a renormalization scale of $\zeta=19$ GeV. This equation arises from considering the factorization of the Schwinger phase in the quark propagator, along with rainbow truncation and the assumption that the gluon propagator is unaffected by the magnetic field.

\section{Numerical results}\label{sec:results}

Given the structures of the inverse propagator and propagator in Eqs.~\eqref{eq:inversestructures} and \eqref{eq:prostructures}, respectively, we can use the gap equation (Eq.~\eqref{eq:gapeqmom}) to determine the propagator. By projecting the gap equation onto different Dirac components, we derive a set of scalar equations, Eq.~\eqref{EqsAtoE},  as detailed in \cref{appendix:gapeq}. For numerical analysis, we transform the system to Euclidean space via a Wick rotation, where the dressing functions depend on the longitudinal momentum $p_l^2 = p_0^2 + p_z^2$ and the transverse momentum $p_t^2 = p_x^2 + p_y^2$.

Additionally, we aim to investigate the behavior of the quark propagator for different quark flavors under varying magnetic fields. The strange quark serves as an excellent reference when compared to the more commonly studied up and down quarks. As noted in Refs.~\cite{Chen:2012qr,Ding:2015rkn}, the strange quark exhibits unique properties within hadrons. In this study, we examine the behavior of up and down quarks in relation to the strange quark under the influence of an external magnetic field. To proceed with numerical computations, it is essential to specify the current quark masses for each flavor. For light quarks - up ($u$), down ($d$) and strange ($s$) - assuming isospin symmetry where  $m_l:=m_u=m_d$, we adopt values from Ref.~\cite{Qin:2019hgk}:
\begin{align}
    m_l^{\zeta=19\,\text{GeV}} = 3.3\,\text{MeV}\,, \quad m_s^{\zeta=19\,\text{GeV}} = 74.6\,\text{MeV}\,.
\end{align}
These values reproduce the empirical masses and leptonic decay constants of pions and kaons.

With these inputs established, we are ready to perform numerical computations of Eq.~\eqref{EqsAtoE} to determine the values of $\hat{A} $ to $ \hat{E}$. Using relations from Eq.~\eqref{eq:AEtoST}, these results can be transformed into $\hat{\mathbb{S}}$, $\hat{\mathbb{V}}_\parallel$, $\hat{\mathbb{V}}_\perp$, $\hat{\mathbb{A}}$ and $\hat{\mathbb{T}}$, thereby establishing a clearer connection to those known in the absence of an external magnetic field.

\subsection{Zero Magnetic Field}

In the absence of a magnetic field ($h=0$), the dressing functions satisfy the relations in Eqs.~\eqref{eq:AEh0}, and \eqref{eq:STh0}. Consequently, the only non-trivial dressing functions are $\hat{\mathbb{S}}$ and $\hat{\mathbb{V}}$, allowing us to define a unique quark mass function $\hat{\mathbb{M}}^{\text{eff}}$ as shown in Eq.~\eqref{eq:Mh0}. The numerical values for $\hat{\mathbb{S}}_{h=0}(p_l^2,p_t^2)$ and $\hat{\mathbb{V}}_{h=0}(p_l^2,p_t^2)$ and $\hat{\mathbb{M}}^{\text{eff}}_{h=0}(p_l^2,p_t^2)$ are obtained.

\begin{table}[t]
\begin{tabularx}{1.0\columnwidth}{ | >{\centering\arraybackslash}X | >{\centering\arraybackslash}X | >{\centering\arraybackslash}X | >{\centering\arraybackslash}X | >{\centering\arraybackslash}X | >{\centering\arraybackslash}X | >{\centering\arraybackslash}X | } 
 \hline
  &  $\hat{\mathbb{S}}^{u,d}$ & $\hat{\mathbb{V}}^{u,d}_{\parallel,\perp}$& $\hat{\mathbb{M}}^{u,d}_{\parallel,\perp}$ &$\hat{\mathbb{S}}^{s}$ & $\hat{\mathbb{V}}^{s}_{\parallel,\perp}$  & $\hat{\mathbb{M}}^{s}_{\parallel,\perp}$\\ 
 \hline
$\omega=0.4$ & 1.26 & 2.01 & 0.63 &1.46&1.81&0.81 \\ 
$\omega=0.5$  & 0.88  & 1.65 & 0.54&1.16&1.60&0.72\\ 
$\omega=0.6$  & 0.59 & 1.35 & 0.44&0.91&1.41&0.65\\
 \hline
\end{tabularx}
\caption{Dressing functions of quark propagator for different flavors - $\hat{\mathbb{S}}$, $\hat{\mathbb{V}}$ - and the effective mass function $\hat{\mathbb{M}}^{\text{eff}}$, evaluated at $p_l^2=p_t^2=0$ in the absence of a magnetic field. The units for $\hat{\mathbb{S}}$ and $\hat{\mathbb{M}}^{\text{eff}}$ are in GeV, and $\hat{\mathbb{V}}$ is dimensionless.}
\label{table:h0plpt0}
\end{table}

At zero momentum ($p_l^2=p_t^2=0$), the values of these dressing functions are summarized in Table~\ref{table:h0plpt0}. In particular, it is confirmed that $\hat{\mathbb{V}}_\parallel=\hat{\mathbb{V}}_\perp$ when no magnetic field is present. Furthermore, $\hat{\mathbb{M}}^{\text{eff}}_{h=0}(0,0)$ corresponds to the quark mass function at zero momentum, with the computed values for different quark flavors aligning with other estimates.

\begin{figure}[t]
\includegraphics[width=0.99\columnwidth]{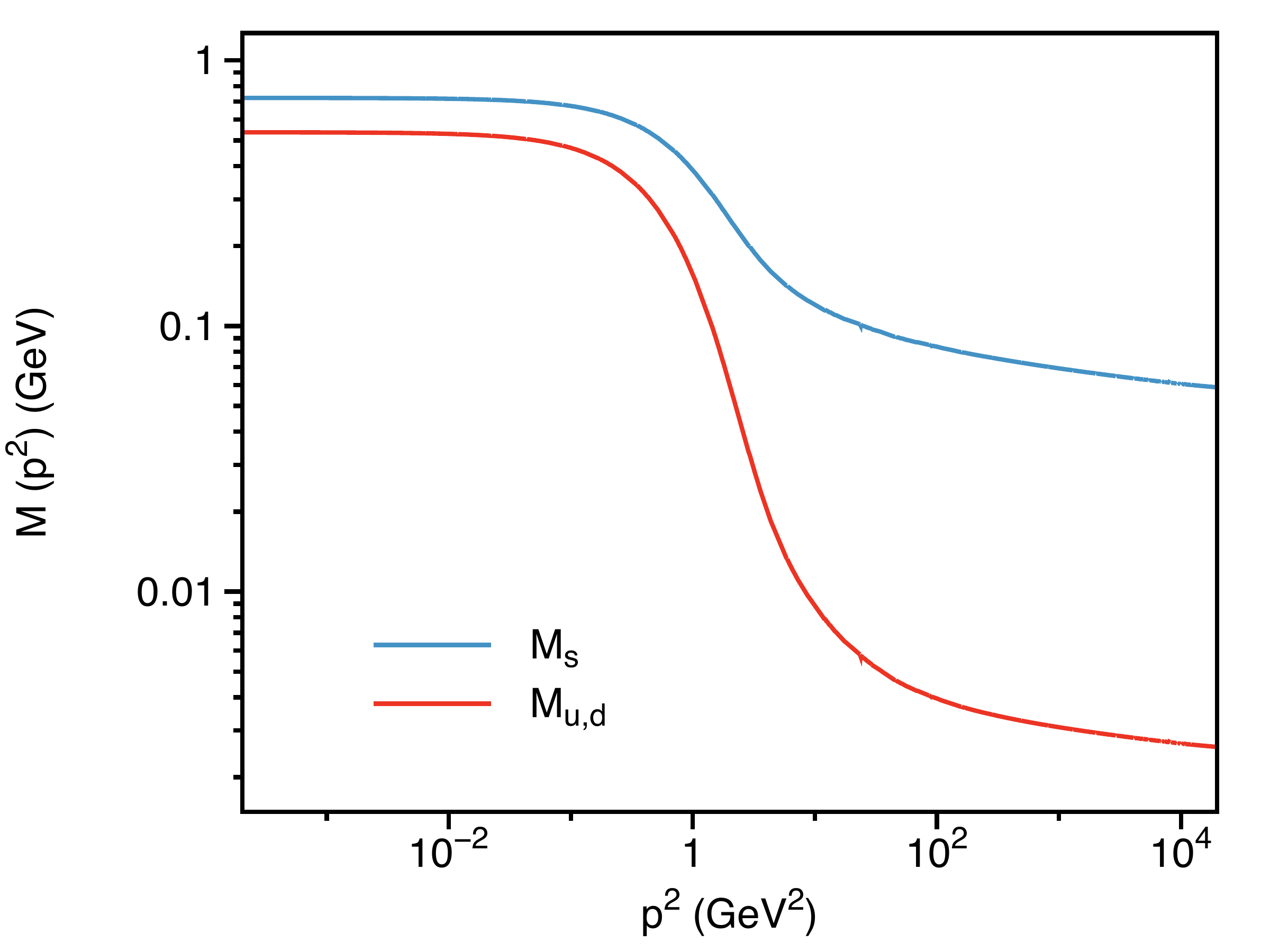}
\caption{The mass function of the quark propagator, denoted as $\hat{\mathbb{M}}^{\text{eff}}$, evaluated at $p^2=p_l^2+p_t^2$ in the absence of a magnetic field.}
\label{fig:Mp}
\end{figure}

At nonzero momentum ($p_l^2\neq 0$, and $p_t^2\neq 0$), we present the results for $\omega = 0.5$ GeV as an illustrative example. The numerical results for $\hat{\mathbb{M}}^{\text{eff}}_{h=0}(p^2)$, where $p^2 = p_l^2 + p_t^2$, are shown in Fig.~\ref{fig:Mp}. The figure reveals that the mass function exhibits characteristic behavior: it attains non-zero values in the infrared region and approaches the current quark mass in the ultraviolet limit, indicating that the quark propagator resembles that of a free quark in the ultraviolet region. Additionally, the figure highlights distinctions between the mass functions of various quark flavors. Up and down quarks experience significant gluon dressing in the infrared region, which is central to dynamical chiral symmetry breaking, and plays a crucial role in the emergent mass of hadrons (see, for example, Ref.~\cite{Ding:2022ows}). In contrast, the strange quark undergoes less extensive dressing, with explicit chiral symmetry breaking driven by the current quark mass, playing a comparative role.

\subsection{Weak Magnetic Field}

In the presence of a non-zero magnetic field ($h\neq 0$), Eqs.~\eqref{eq:AEh0}, ~\eqref{eq:STh0} no longer apply, resulting in all dressing functions - $\hat{\mathbb{S}}$, $\hat{\mathbb{V}}_\parallel$, $\hat{\mathbb{V}}_\perp$, $\hat{\mathbb{A}}$ and $\hat{\mathbb{T}}$ - becoming nonzero..

\subsubsection{ \texorpdfstring{$\hat{\mathbb{S}}$, $\hat{\mathbb{V}}_\parallel$ and $\hat{\mathbb{V}}_\perp$}{Lg}}

We begin by analyzing the dressing functions associated with the scalar and vector Dirac structures: $\hat{\mathbb{S}}$, $\hat{\mathbb{V}}_\parallel$, and $\hat{\mathbb{V}}_\perp$. These functions are intrinsically linked to the characteristics of the quark propagator in vacuum and encapsulate the average-spin properties of the quark, providing insights into mass and momentum dynamics in both longitudinal and transverse directions relative to the magnetic field.

\begin{figure}[htbp]
\includegraphics[width=0.92\columnwidth]{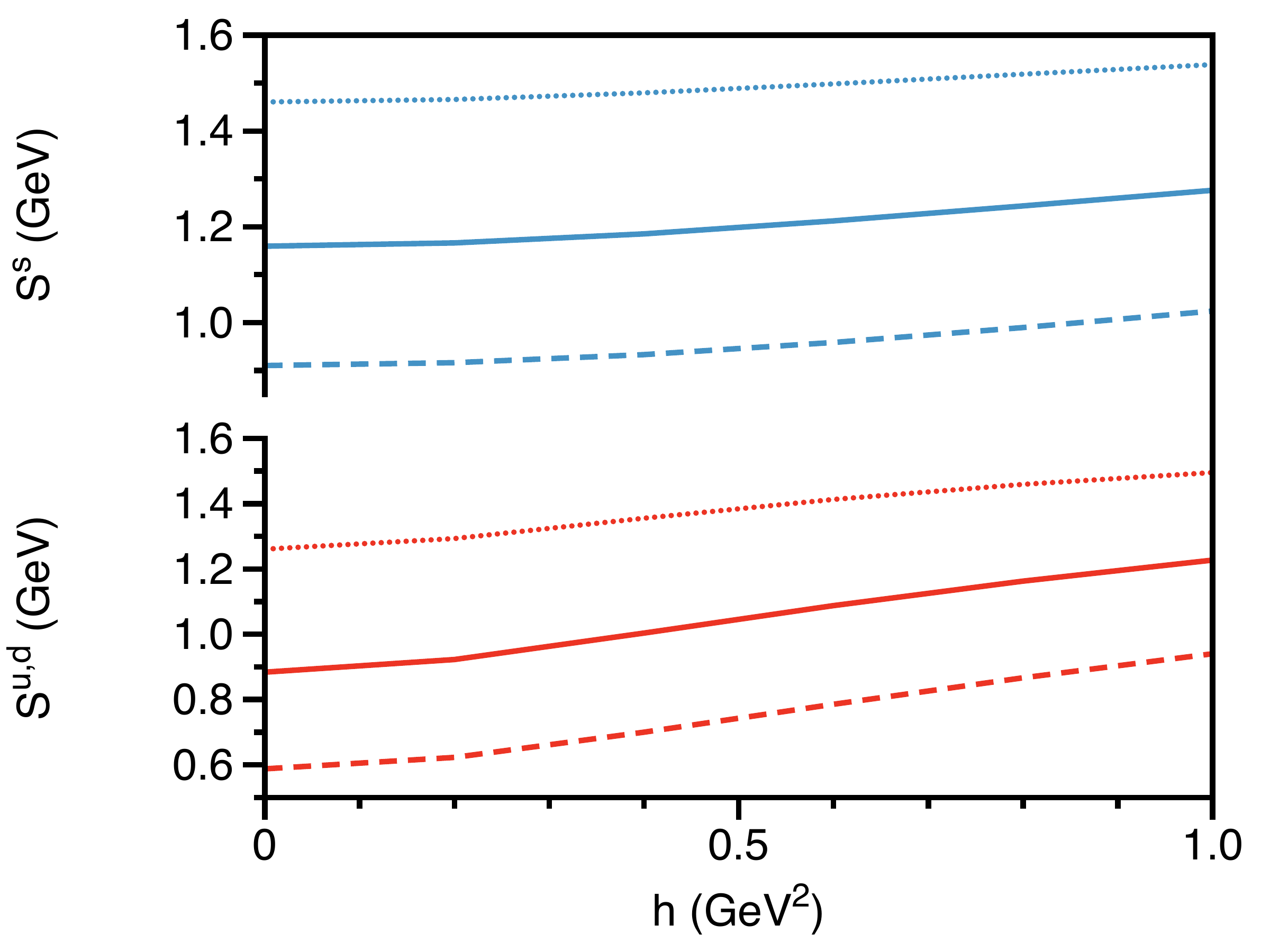}
\includegraphics[width=0.92\columnwidth]{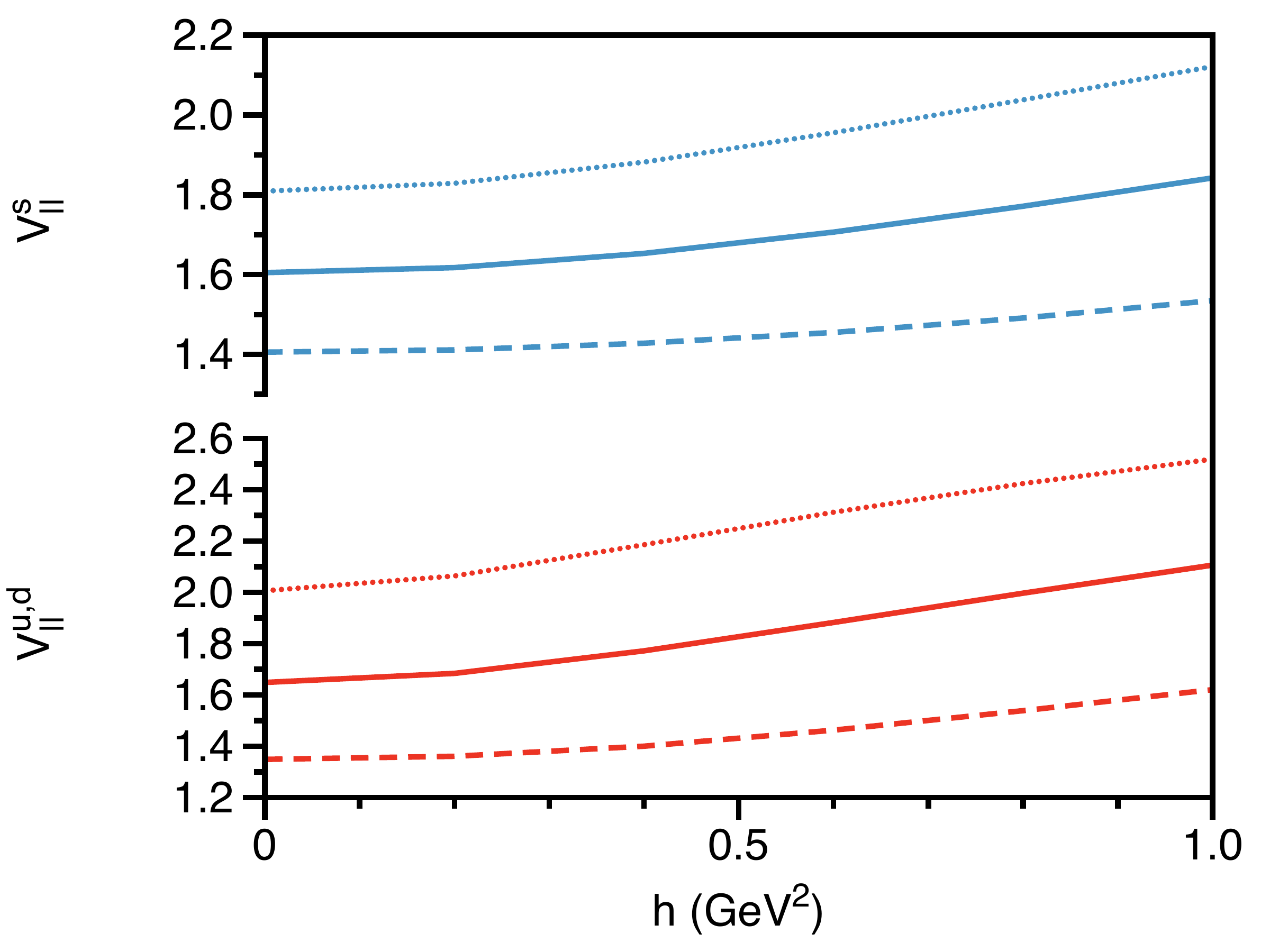}
\includegraphics[width=0.92\columnwidth]{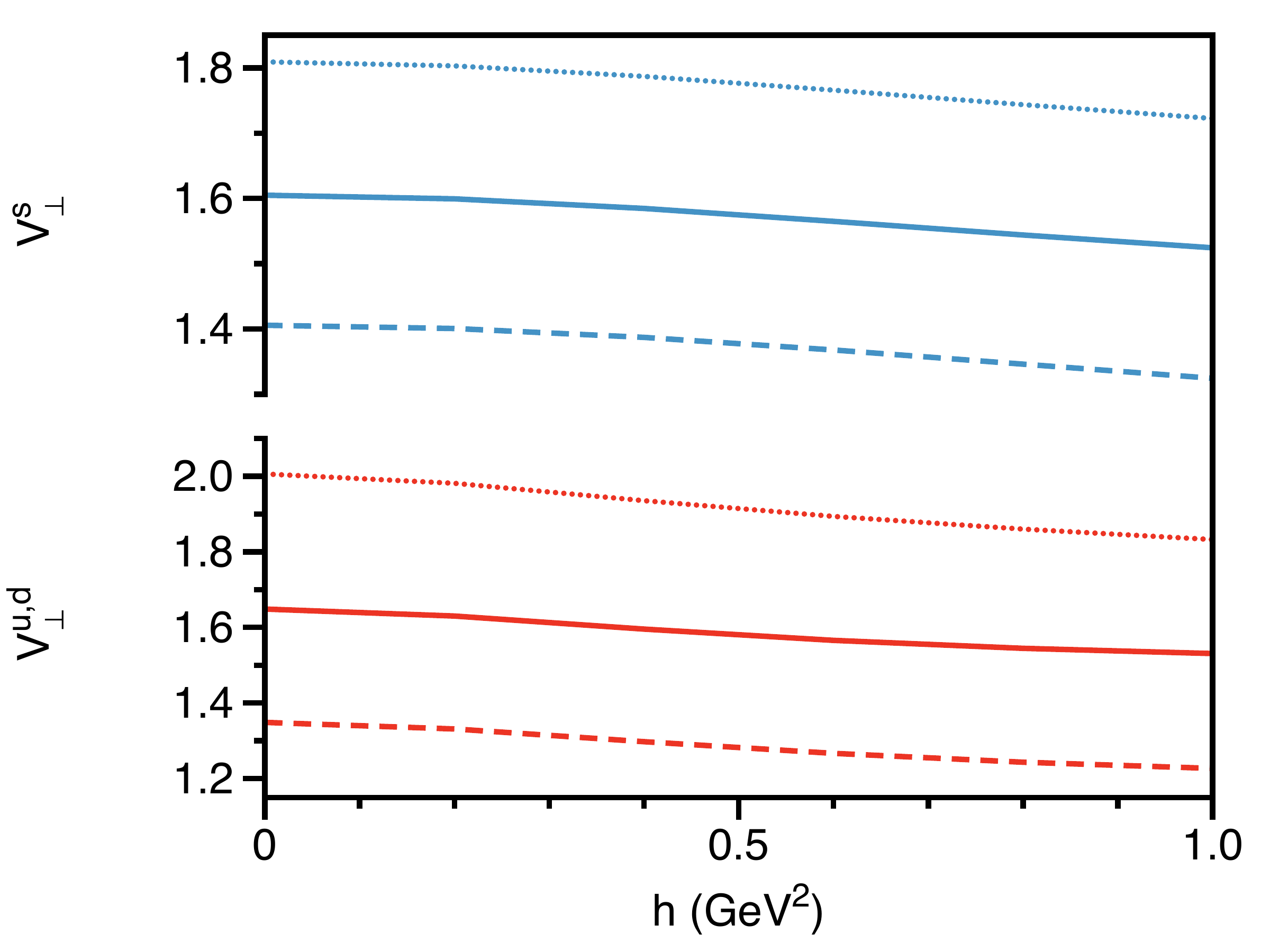}
\caption{Magnetic field dependence of the dressing functions for the inverse up/down and strange quark propagators - $\hat{\mathbb{S}}$, $\hat{\mathbb{V}}_\parallel$, and $\hat{\mathbb{V}}_\perp$ - evaluated at $p_l^2 = p_t^2 = 0$. Curves: Dotted for $\omega = 0.4$ GeV, Solid for $\omega = 0.5$ GeV, and Dashed for $\omega = 0.6$ GeV.}
\label{fig:svplpt00}
\end{figure}

At zero momentum, the discrete data points in Table \ref{table:h0plpt0} transform into magnetic field-dependent curves, as shown in Fig. \ref{fig:svplpt00}. The figure indicates that the overall behavior is similar for both up/down and strange quarks; specifically, as the magnetic field increases, the dressing functions consistently increase or decrease across the entire range. The scalar dressing function $\hat{\mathbb{S}}$ increases notably with the magnetic field, consistent with magnetic catalysis. In contrast, the vector dressing functions exhibit notable differences, with $\hat{\mathbb{V}}_\parallel \neq \hat{\mathbb{V}}_\perp$, indicating anisotropic gluon dressing of longitudinal momentum $\slashed{p}_\parallel$ and transverse momentum $\slashed{p}_\perp$ in the presence of a magnetic field. This anisotropy reflects how the magnetic field differentially influences the gluon dressing in parallel versus perpendicular directions. As the magnetic field increases, the dressing function parallel to the field, $\hat{\mathbb{V}}_\parallel$, increases, while the dressing function perpendicular to the field, $\hat{\mathbb{V}}_\perp$, decreases. Furthermore, variations in the interaction width of the input gluon propagator parameter $\omega$ do not alter these qualitative conclusions.

\begin{figure}[t]
\includegraphics[width=0.99\columnwidth]{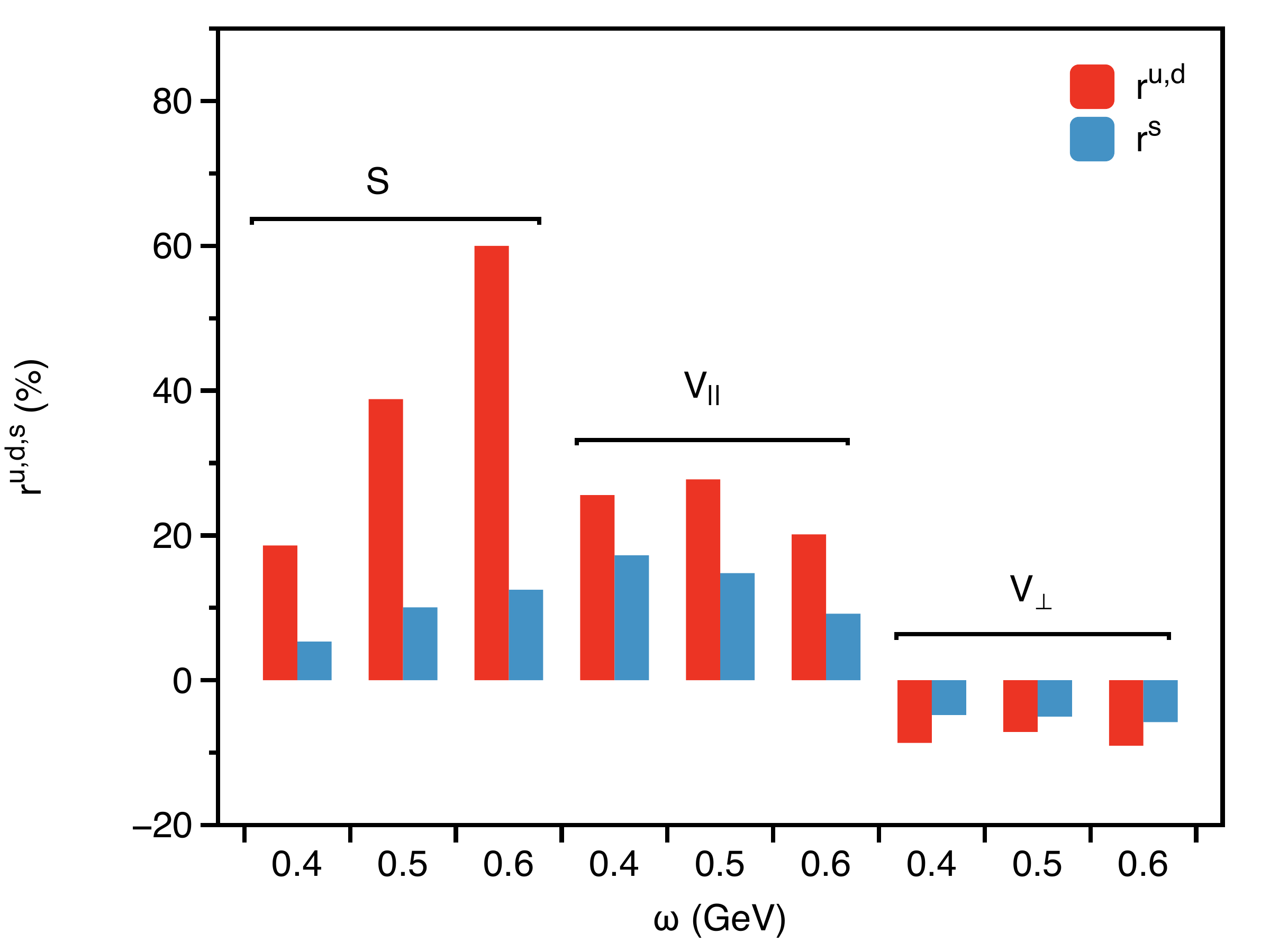}
\caption{Percentage variation in the dressing functions of the quark propagator -  $\hat{\mathbb{S}}$, $\hat{\mathbb{V}}_{\parallel}$, and $\hat{\mathbb{V}}_{\perp}$ - evaluated at $p_l^2 = p_t^2 = 0$, comparing cases with $h=0$ and $h=1$ GeV$^2$, as defined in Eq.~\eqref{eq:varyrate}.}
\label{fig:svpecnet}
\end{figure}

To evaluate the change in these dressing functions, we compute the rate of change defined by:
\begin{align}\label{eq:varyrate}
	r^f_{\hat{\mathbb{F}}} := \frac{\hat{\mathbb{F}}^{f}_{h=1.0\,\text{GeV}^2}(0,0) - \hat{\mathbb{F}}^f_{h=0}(0,0)}{\hat{\mathbb{F}}^f_{h=0}(0,0)},
\end{align}
where $\hat{\mathbb{F}} = \hat{\mathbb{S}}, \hat{\mathbb{V}}_\parallel, \hat{\mathbb{V}}_\perp$ and $f = u, d, s$.  The results are summarized in Fig.~\ref{fig:svpecnet}. The figure indicates that the magnetic field induces a more pronounced change in the dressing functions of the up/down quark compared to that of the strange quark, reflecting a stronger response of the up/down quark to the magnetic field. Specifically, the percentage increases in the scalar dressing function $\hat{\mathbb{S}}$ are less than $60\%$ for the up/down quark and approximately $15\%$ for the strange quark when comparing cases with $h=0$ and $h=1$ GeV$^2$. As the magnetic field strength increases, $\hat{\mathbb{V}}_\parallel$ increases by about $30\%$ for the up/down quark and $20\%$ for the strange quark, while $\hat{\mathbb{V}}_\perp$ decreases by approximately $10\%$ and $6\%$, respectively.

Since $\hat{\mathbb{V}}_\parallel\neq\hat{\mathbb{V}}_\perp$ in the presence of a non-zero magnetic field, we define two distinct mass functions according to Eq.~\eqref{eq:Mhnonzero}, resulting in $\hat{\mathbb{M}}^{\text{eff}}_\parallel\neq\hat{\mathbb{M}}^{\text{eff}}_\perp$.  In their concluding remarks, Watson et al.~\cite{Watson:2013ghq} note the separation of mass functions as the magnetic field strength increases. This study builds upon their findings, offering a detailed exploration of these effects across various quark flavors.

\begin{figure}[t]
\includegraphics[width=0.92\columnwidth]{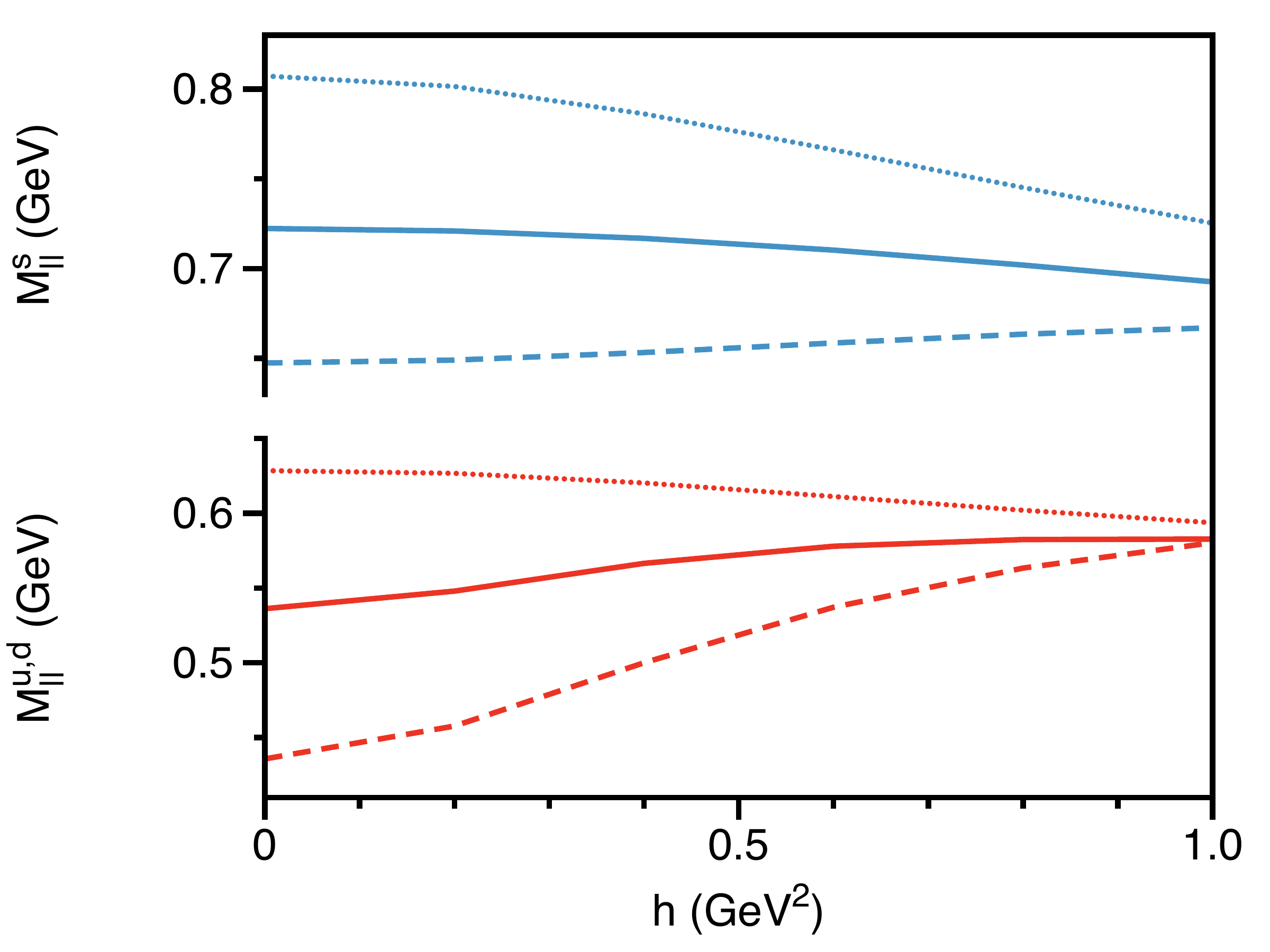}
\includegraphics[width=0.92\columnwidth]{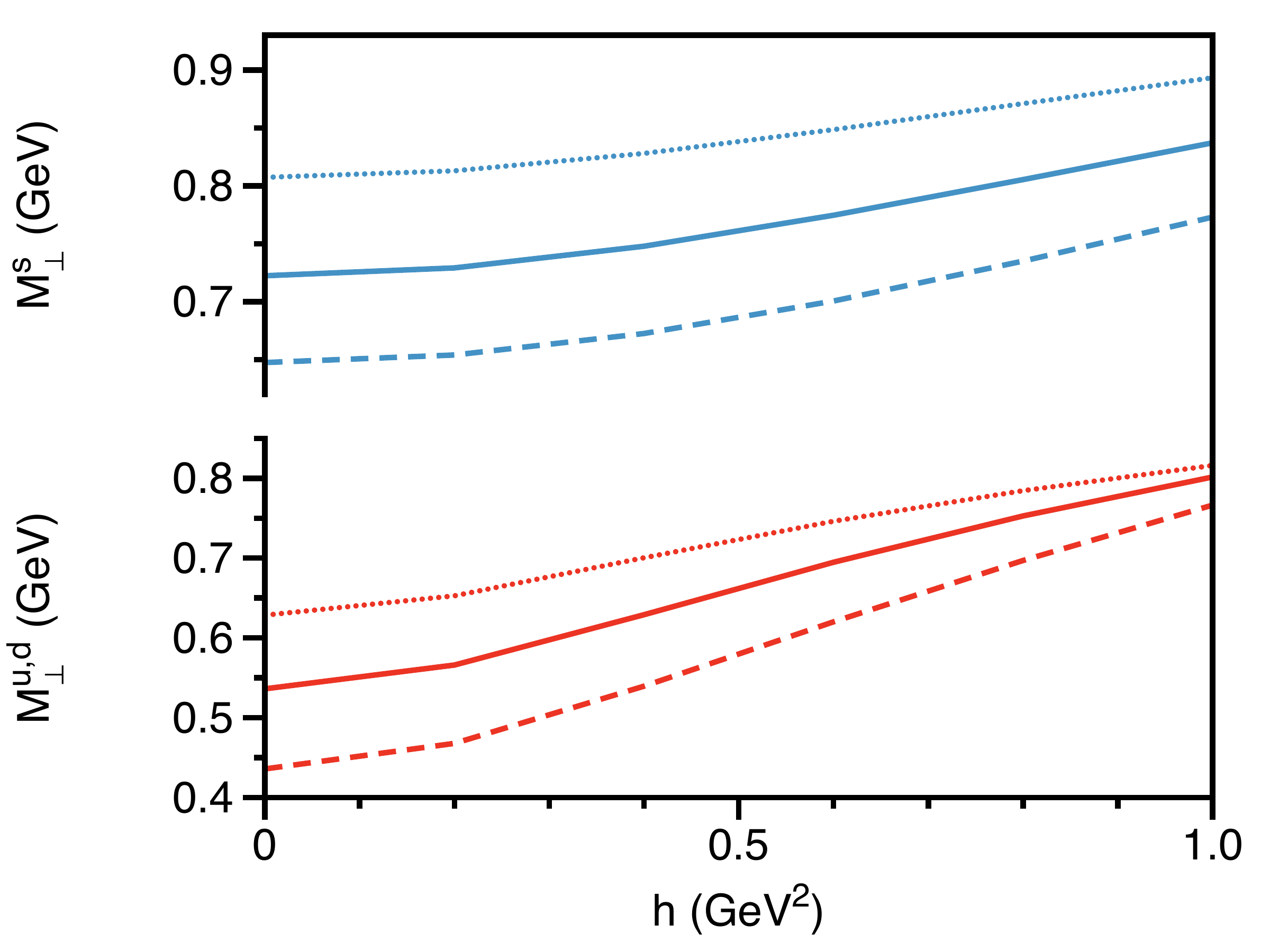}
\includegraphics[width=0.92\columnwidth]{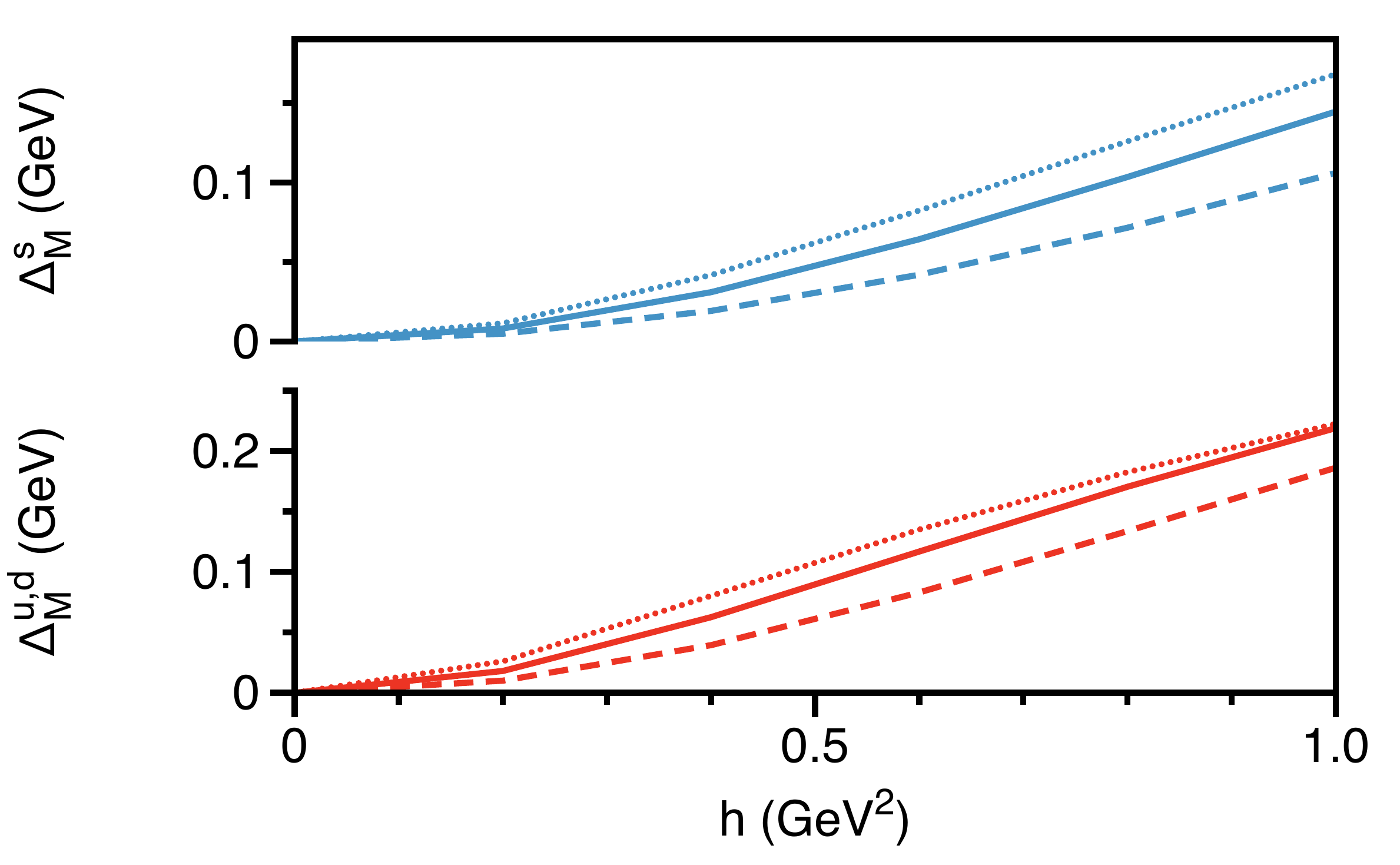}
\caption{Magnetic field dependence of the effective mass functions for the up/down and strange quarks - $\hat{\mathbb{M}}^{\text{eff}}_\parallel$, $\hat{\mathbb{M}}^{\text{eff}}_\perp$ and their difference $\Delta_M=\hat{\mathbb{M}}^{\text{eff}}_\perp-\hat{\mathbb{M}}^{\text{eff}}_\parallel$ -  evaluated at $p_l^2 = p_t^2 = 0$. Curves: Dotted for $\omega = 0.4$ GeV, Solid for $\omega = 0.5$ GeV, and Dashed for $\omega = 0.6$ GeV.}
\label{fig:mplpt00}
\end{figure}

The numerical results for the two mass functions are illustrated in Fig.~\ref{fig:mplpt00}. As the scalar dressing function $\hat{\mathbb{S}}$ increases and the transverse vector dressing function $\hat{\mathbb{V}}_\perp$ decreases with increasing magnetic field, it follows that the transverse effective mass function $\hat{\mathbb{M}}^{\text{eff}}_{\perp}$ also increases. In contrast, the behavior of the longitudinal effective mass function $\hat{\mathbb{M}}^{\text{eff}}_{\parallel}$ is more complex due to the concurrent increase in $\hat{\mathbb{V}}_\parallel$, which is influenced by the quark gluon interaction width $\omega$. Notably, using the commonly adopted interaction width of $\omega=0.5$ GeV, the variation in $\hat{\mathbb{M}}^{\text{eff}}_{\parallel}$ is relatively modest. Nevertheless, in all quark flavors considered, the inequality $\hat{\mathbb{V}}_\parallel>\hat{\mathbb{V}}_\perp$ holds consistently, resulting in the effective mass $\hat{\mathbb{M}}^{\text{eff}}_{\perp}>\hat{\mathbb{M}}^{\text{eff}}_{\parallel}$.

\begin{figure}[t]
\includegraphics[width=0.99\columnwidth]{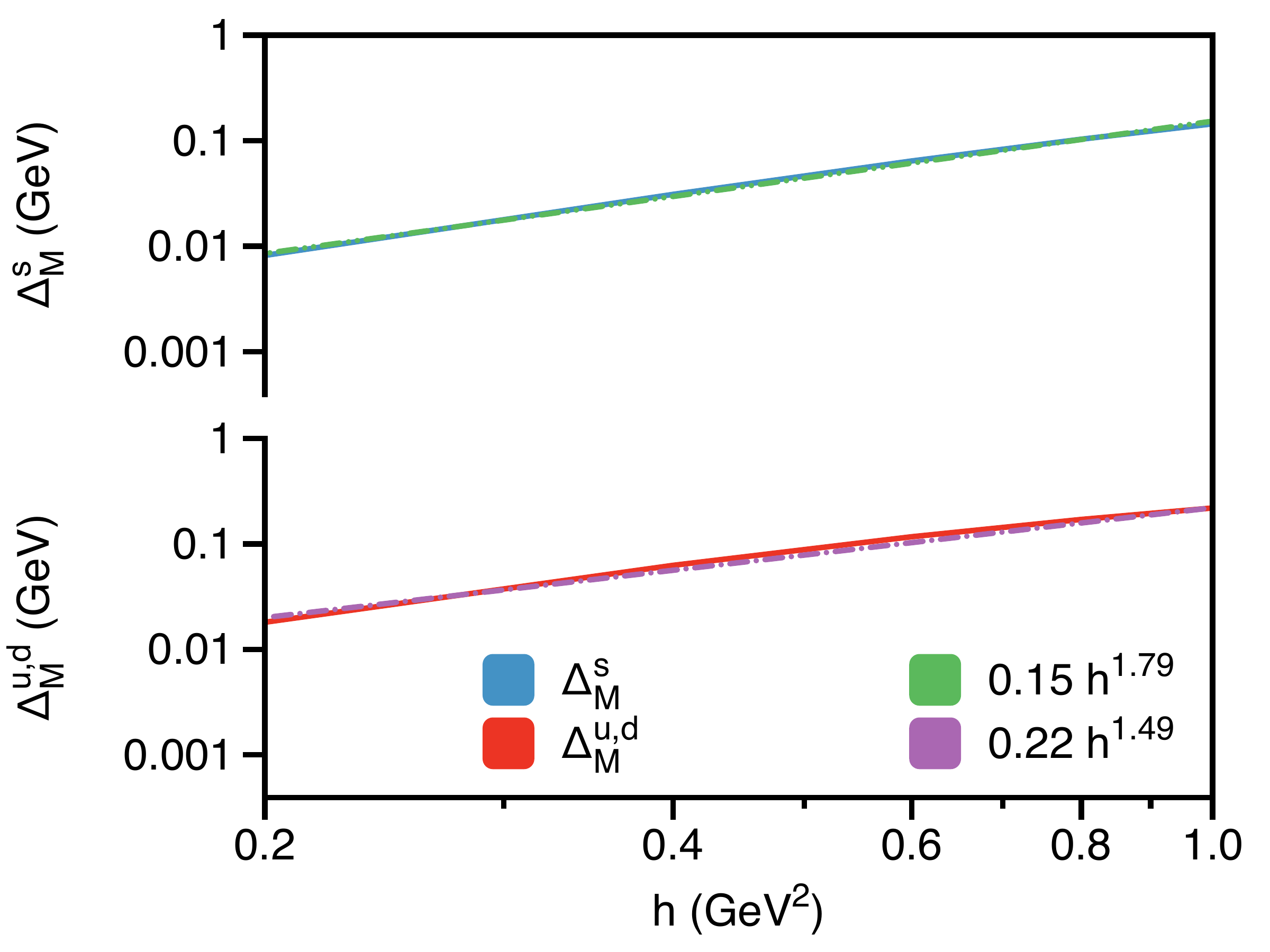}
\caption{Logarithmic plot illustrating ${\hat{\mathbb{M}}^{\text{eff}}_\perp}(0,0)-{\hat{\mathbb{M}}^{\text{eff}}_\parallel}(0,0)$.}
\label{fig:masssplitlog00}
\end{figure}

To quantify the difference between the transverse and longitudinal effective masses, we define $\Delta_M = \hat{\mathbb{M}}^{\text{eff}}_\perp - \hat{\mathbb{M}}^{\text{eff}}_\parallel$ and examine its variation with respect to the magnetic field, as shown in Fig.~\ref{fig:mplpt00}. The figure clearly demonstrates that the mass splitting $\Delta_M$ increases with the magnetic field strength across all quark flavors, with the strange quark exhibiting smaller mass splittings than the up/down quark. Specifically, comparing cases with $h=0$ and $h=1$ GeV$^2$, yields a mass difference of $\Delta_M^{u,d} = 0.20(2)$ GeV for the up/down quark and $\Delta_M^s = 0.14(3)$ GeV for the strange quark (error estimates come from different $\omega$). Moreover, Fig.~\ref{fig:mplpt00} suggests a scaling behavior for $\Delta_M(h)$. To explore this further, we present a logarithmic graph of $\Delta_M$ in Fig.~\ref{fig:masssplitlog00} (with $\omega = 0.5$ GeV), which reveals a linear relationship between $\log\Delta_M$ and $\log h$. This observation suggests that the numerical results can be approximately described by a power law:
\begin{align}
\Delta_M^{u,d}(h) = 0.22\, h^{1.49}\,, \quad \quad \Delta_M^{s}(h) = 0.15 \,h^{1.79}.	
\end{align}
The scaling exponent of the mass splitting induced by the magnetic field appears to gradually approach $2$ as one transitions from light to heavy quarks.

At nonzero momentum, the effective mass functions we derive depend primarily on the squares of the longitudinal and transverse momenta, $p_l^2$, and $p_t^2$. Notably, we observe an almost symmetric dependence between these two momenta, despite the absence of explicit symmetry constraints. This near-symmetry may arise from the fact that while the quark propagator explicitly differentiates between longitudinal and transverse directions through its Dirac structures, the gluon propagator employed in our analysis does not. Consequently, although the magnetic field induces different gluon dressing effects in the longitudinal and transverse directions, the dependence of the dressing functions - and thus the effective mass functions - on $p_l^2$, and $p_t^2$ remains nearly symmetric. Future studies could refine this analysis by incorporating a gluon propagator with explicit longitudinal and transverse components.

\begin{figure}[t]
\includegraphics[width=0.99\columnwidth]{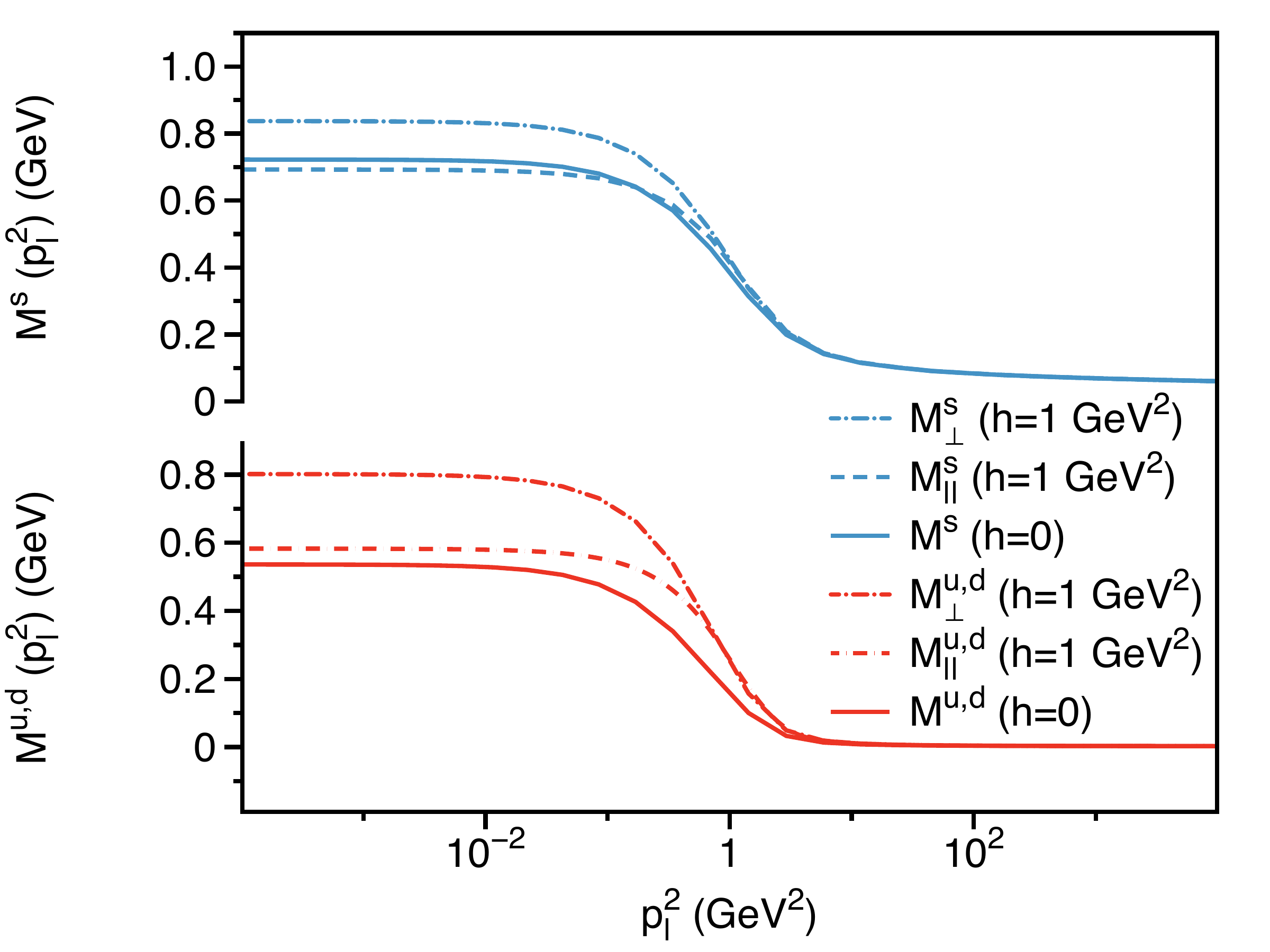}
\caption{Comparison of the quark mass functions $\hat{\mathbb{M}}^{u,d,s}_{\parallel,\perp}(p_l^2, 0)$ at $h=1.0$ GeV$^2$ versus their values at zero magnetic field ($h=0$).}
\label{fig:mplpth10}
\end{figure}

Given this observed near-symmetry, we focus solely on the dependence of the effective mass functions on the longitudinal momentum square, $ p_l^2 $. To illustrate the variation in momentum dependence of the effective mass function under different magnetic field strengths, we compare the mass functions at zero magnetic field ($ h = 0 $), as shown in Fig. \ref{fig:Mp}, with those at a non-zero magnetic field, specifically $ h = 1.0 \, \text{GeV}^2 $, depicted in Fig. \ref{fig:mplpth10}. For this analysis, we use $ \omega = 0.5 \, \text{GeV} $ as a representative example. The figure demonstrates that, while the overall qualitative behavior of the mass function remains largely unchanged, variations in the infrared region are noted. Consequently, we have numerically determined the mass functions defined in Eq.~\eqref{eq:Mhnonzero} across the entire momentum range for up/down and strange quarks. This complete momentum-dependent mass function is crucial for studying physical observables influenced by magnetic fields, as these quantities inherently involve integrals of the momentum-dependent mass function over the full momentum range.

\subsubsection{\texorpdfstring{$\hat{\mathbb{A}}$ and $\hat{\mathbb{T}}$}{Lg}}

The axial-vector and tensor structures are new Dirac structures that appear in the quark propagator only in the presence of a magnetic field, where they become non-zero (i.e., $\hat{\mathbb{A}}\neq 0$ and $\hat{\mathbb{T}}\neq 0$). We present results for $h\hat{\mathbb{A}}$ and $2h\hat{\mathbb{T}}$, as these combinations are the actual dressing functions that appear in the quark propagator, as shown in Eq.~\eqref{eq:16diracbasis}. As previously discussed, these two terms capture the intricate interplay between quark spin and the external magnetic field, highlighting the resulting asymmetry between spin-up and spin-down states. At each Landau level, both terms contribute to the Zeeman splitting of the energy levels associated with different spin states.

\begin{figure}[t]
\includegraphics[width=0.99\columnwidth]{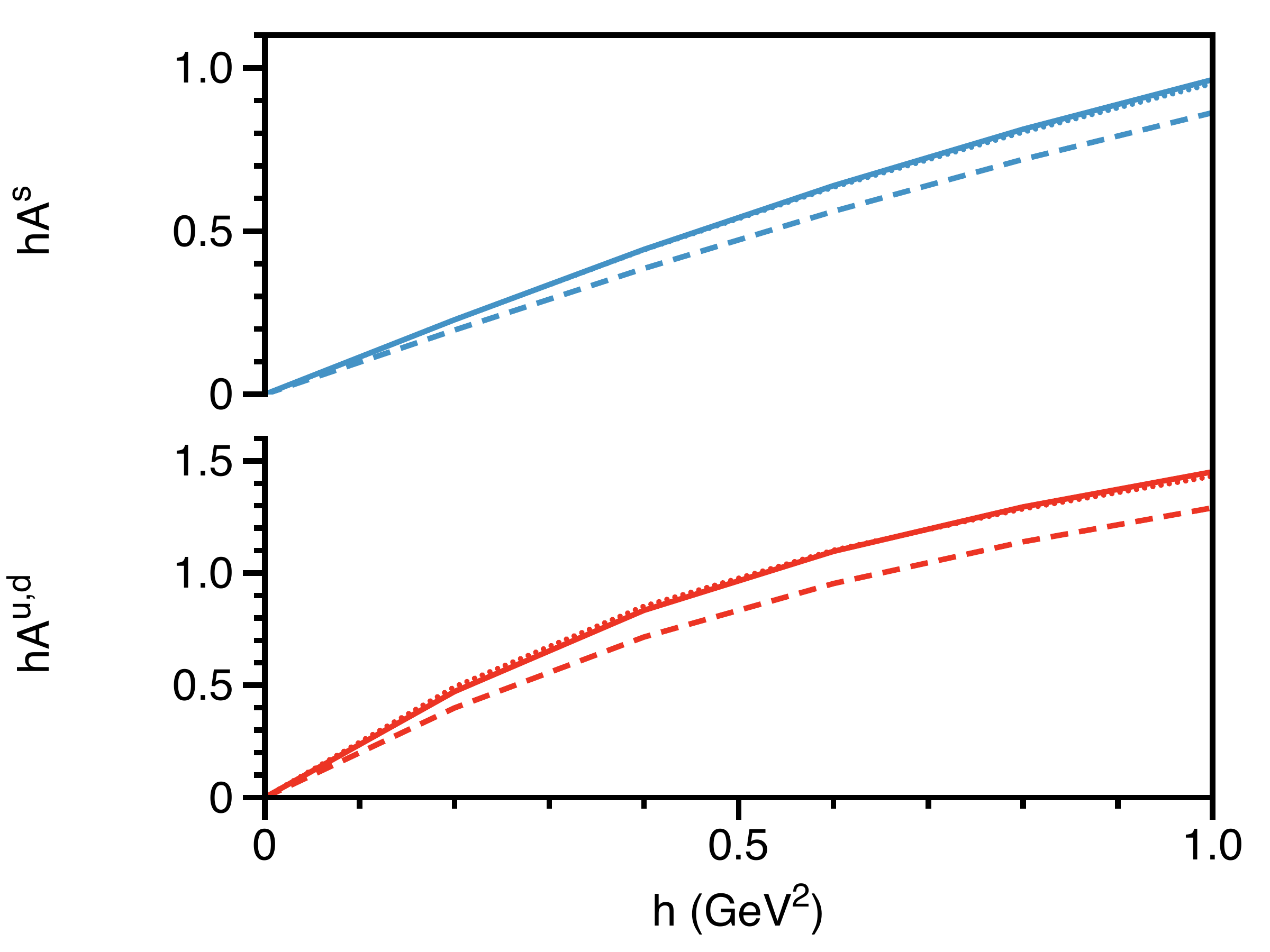}
\includegraphics[width=0.99\columnwidth]{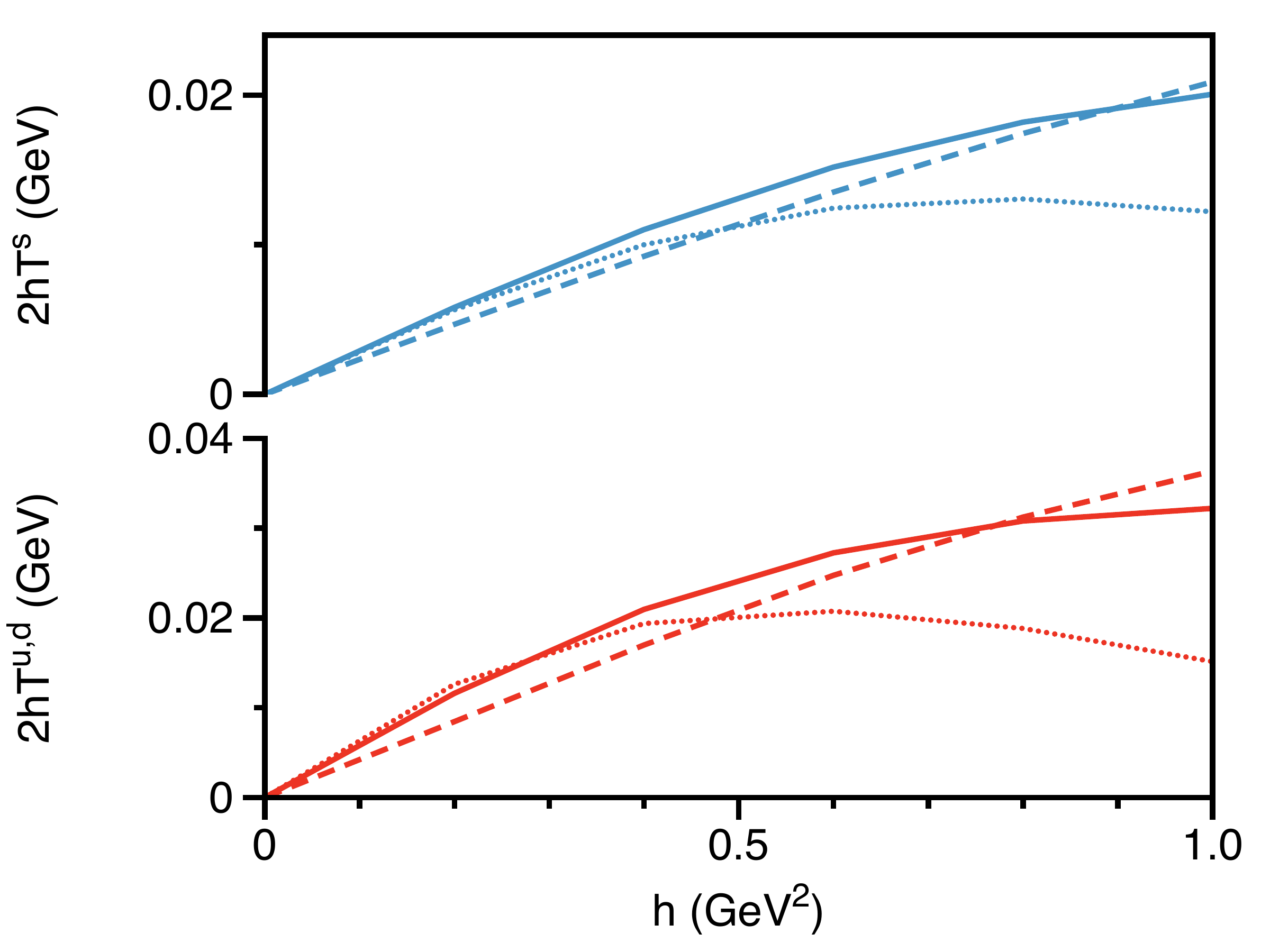}
\caption{Magnetic field dependence of the dressing functions, represented as $h\hat{\mathbb{A}}$, and $2h\hat{\mathbb{T}}$, evaluated at $p_l^2 = p_t^2 = 0$. Curves: Dotted for $\omega = 0.4$ GeV, Solid for $\omega = 0.5$ GeV, and Dashed for $\omega = 0.6$ GeV.}
\label{fig:atplpt00}
\end{figure}

At zero momentum, the dependence of $h\hat{\mathbb{A}}(0,0)$ and $2h\hat{\mathbb{T}}(0,0)$ on the magnetic field is illustrated in Fig.~\ref{fig:atplpt00}. Both quantities are zero when $ h = 0 $. The dressing function associated with the axial-vector structure, $ h\hat{\mathbb{A}} $ (upper panel of Fig.~\ref{fig:atplpt00}), shows a noticeable increase with the magnetic field for both up/down and strange quarks. Although this increase is substantial, its rate decreases for heavier quarks, although their contribution remains significant (on the order of $1$). The width of the quark-gluon interaction $\omega$ has only a minor influence on this increase, showing no evident dependence.

The dressing function associated with the tensor structure, $2h\hat{\mathbb{T}}(0,0)$ (lower panel of Fig.~\ref{fig:atplpt00}),  does not always increase monotonically. In the low $h$ region, it exhibits a linear pattern, consistent with Schwinger's prediction that the leading contribution proportional to $\sigma_{\mu\nu}$ is linear in the field. In the high $h$ region, its behavior depends on the width of the quark-gluon interaction $\omega$, reflecting that tensor structures are particularly sensitive to interaction details. When $\omega$ is small, $2h\hat{\mathbb{T}}$ shows a downward trend in strong magnetic fields. Nevertheless, heavier quarks continue to show smaller variations in $2h\hat{\mathbb{T}}$. Furthermore, its magnitude is approximately $0.01$, which is smaller than all other dressing functions. The tensor terms encapsulate the mass asymmetry between spin-up and spin-down states: the spin-up quark exhibits a reduced effective mass due to lower energy, while the spin-down quark experiences an increased effective mass due to higher energy. The mass differences between these states grow with increasing magnetic field. However, this effect is minor compared to the variations in effective masses in parallel and perpendicular directions relative to the magnetic field.

\begin{figure}[t]
\includegraphics[width=0.99\columnwidth]{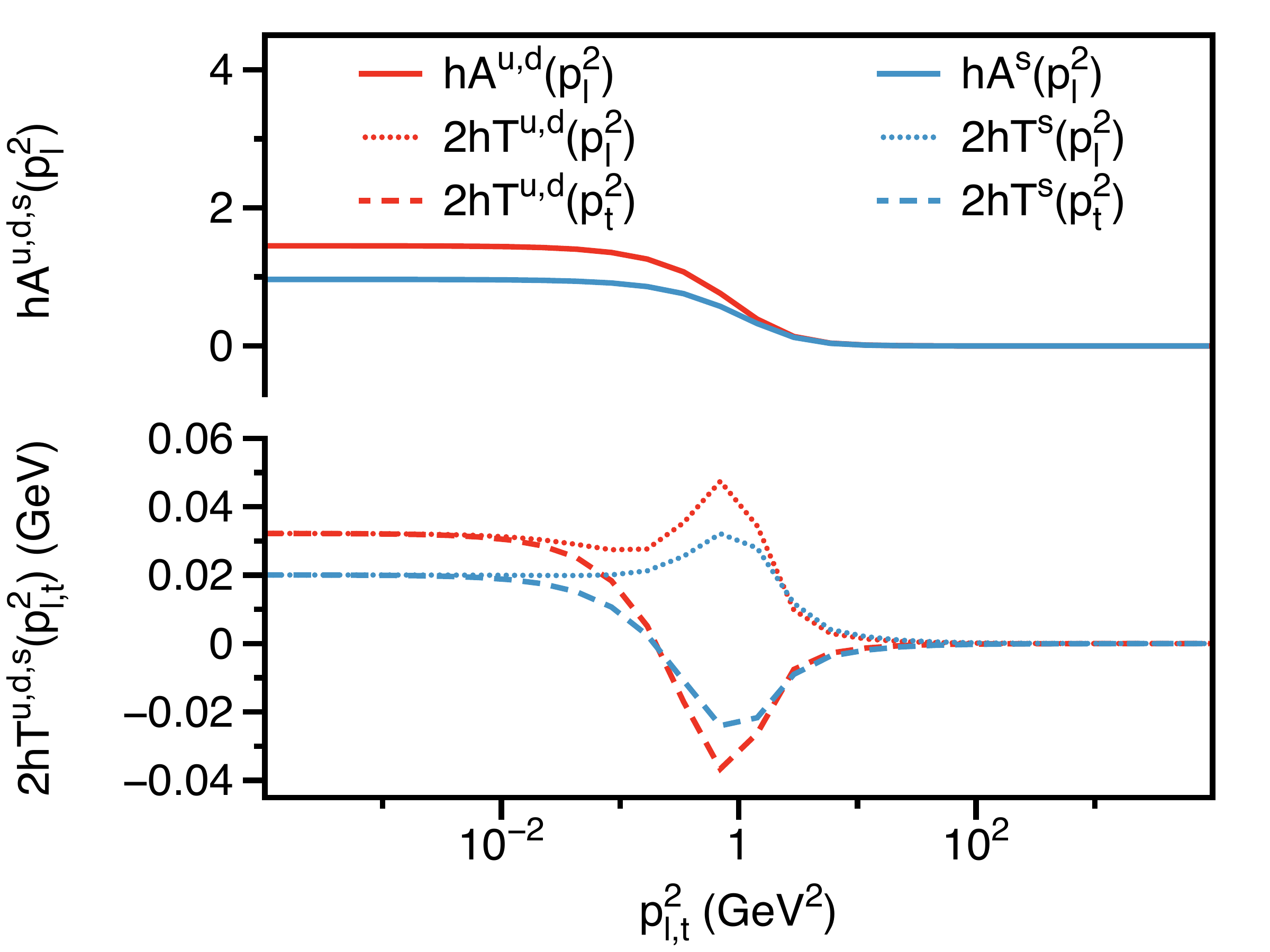}
\caption{Dressing functions of quark propagator: $h\hat{\mathbb{A}}(p_l^2, p_t^2)$, and $2h\hat{\mathbb{T}}(p_l^2, p_t^2)$, evaluated at $h=1.0$ GeV$^2$.}
\label{fig:ATplpth1}
\end{figure}

At nonzero momentum, we use $\omega = 0.5$ GeV as a representative case. The dressing functions at $h = 1.0$ GeV$^2$ are illustrated in Fig.~\ref{fig:ATplpth1}. The behavior of $h\hat{\mathbb{A}}(p_l^2, p_t^2)$ is consistent with previous results, showing symmetry between the longitudinal and transverse momenta ($p_l^2$ and $p_t^2$). Therefore, we present only its dependence on $p_l^2$, which mirrors its dependence on $p_t^2$. As noted previously, $h\hat{\mathbb{A}}(p_l^2, p_t^2)$ remains finite in the infrared and approaches zero in the ultraviolet.

In contrast, the behavior of $2h\hat{\mathbb{T}}(p_l^2, p_t^2)$ deviates significantly. Its dependence on $p_l^2$ and $p_t^2$ differs unexpectedly, despite the fact that the gluon propagator lacks explicit longitudinal and transverse Dirac structures. Nevertheless, we observe that the variation of $2h\hat{\mathbb{T}}(p_l^2, p_t^2)$ is minimal in the infrared. At approximately $p_l^2 \approx p_t^2 \approx 0.1$ GeV$^2$, in the $p_l^2$ direction, $2h\hat{\mathbb{T}}(p_l^2, p_t^2)$ first increases before decreasing; while in the $p_t^2$ direction, it first decreases to negative values before increasing again. Despite these fluctuations, $2h\hat{\mathbb{T}}(p_l^2, p_t^2)$ ultimately tends toward zero in the ultraviolet.

As discussed previously, the axial-vector and tensor terms encapsulate the spin-momentum interaction asymmetry between the spin-up and spin-down states. The increase in these two terms with the magnetic field (excluding the behavior of $ 2h\hat{\mathbb{T}}(p_l^2, p_t^2) $ for a small $\omega$, which decreases in the high $h$ region) can be understood as follows: In the absence of a magnetic field, there is no preferential spin direction, leading to random spin alignment among quarks. The application of a magnetic field aligns more spins with the field direction, enhancing the spin-momentum interaction for spin-up quarks relative to spin-down quarks. This phenomenon increases the system's magnetic moment and reflects the magnetic field's ability to polarize spins, thereby leading to net magnetization - a behavior characteristic of Pauli paramagnetism.

We have numerically obtained and analyzed the fully dressed quark propagators of up/down quarks and strange quarks in the presence of a constant magnetic field. This includes their characteristics at zero momentum, which effectively represents the constituent quark mass, and at across the entire momentum domain. These results serve as fundamental inputs for further investigations into hadron properties, such as quark condensate~\cite{Shushpanov:1997sf,Cohen:2007bt}, potential tensor condensate~\cite{Ioffe:1983ju,Gubler:2018ctz}, quark local magnetic dipole moment~\cite{Buividovich:2009my}, meson and baryon properties, etc., when magnetic field is present.

\section{Summary}\label{sec:summary}

This paper investigates the impact of an external magnetic field on dressed quark propagators for up/down and strange quarks,  focusing on the resulting anisotropic dynamics.



A key finding is the decomposition of the vector part of the propagator into components parallel and perpendicular to the magnetic field, allowing for distinct mass definitions in these directions. Numerical results show that the mass perpendicular to the magnetic field is always greater than that parallel to it, indicating quark propagation anisotropy induced by the magnetic field. The mass difference increases with the magnetic field and follows a power-law behavior, with exponents approximately $1.5$ for up/down quarks and $1.8$ for the strange quark; this difference decreases with increasing quark mass. Additionally, the magnetic field introduces two new terms in the propagator: the axial-vector and tensor terms, which describe interactions between the magnetic field and the spin of quark. Both terms contribute to the Zeeman effect, splitting the energy for spin up and down quarks. The axial-vector term is sizable in the presence of a magnetic field, and its increase suggests quark spin polarization. The tensor term, although smaller in magnitude, shows different momentum dependencies that warrant further investigation. 

The effective masses of quarks exhibit anisotropy in directions parallel and perpendicular to the external magnetic field, consistent with theoretical expectations. This anisotropy arises because quarks experience different constraints in these directions: they are less influenced by the magnetic field along the parallel direction, while their energy levels are quantized into discrete Landau levels in the perpendicular direction, leading to stronger constraints. This behavior results from the magnetic field's impact on quark momentum and spin interactions, producing distinct propagation dynamics. Additionally, this anisotropy varies across quark flavors, with the strange quark showing less pronounced variations compared to the up and down quarks.

Quark propagators in external magnetic fields are crucial for studying a range of QCD phenomena. The investigations here can  shed light on the phenomena that appear in quark matter under extreme conditions, such as  magnetic catalysis/inverse magnetic catalysis, spin polarization and the chiral magnetic effect. As quark propagators represent a one-body problem, they provide the foundational framework for understanding hadron behavior, such as mesons (two-body problems) and baryons (three-body problems). Consequently, the approaches and results discussed here can be extended to explore hadron properties in magnetic fields, offering valuable insights into potential phenomena like $\rho$ meson condensation and neutral pion condensation~\cite{Chernodub:2010qx}.

\begin{acknowledgments}
MD extends gratitude to Craig Roberts and all former colleagues at the Institute of Theoretical Physics of HZDR, particularly Stefan Evans, Burkhard K\"ampfer, Christian Kohlf\"urst, Misha A. Lopez-Lopez, Roland Sauerbrey and Ralf Sch\"utzhold for their valuable discussions. FG is  supported by the National  Science Foundation of China under Grants  No. 12305134. MD appreciates the support received from the Helmholtz-Zentrum Dresden-Rossendorf High Potential Programme.

\end{acknowledgments}


\appendix

\section{Dressed Version of the Dispersion Relation}\label[appendix]{appendix:inverseRitus}

Taking the inverse of the inverse propagator in Eq.~\eqref{eq:ritusdressinv} yields a dressed version of the dispersion relation ($h=q_fB$):
\begin{align}
	\Delta &=0\,,
\end{align}
where
\begin{align}\label{eq:denominator}
	\Delta &=\Delta_1^2-p_\parallel^2\Delta_2^2\,,\notag\\
	\Delta_1 &= p_\parallel^2 \mathcal{A}\mathcal{C}-\mathcal{B}\mathcal{D}-2nh\mathcal{E}^2\,,\notag\\
	\Delta_2 &=\mathcal{A}\mathcal{D}-\mathcal{B}\mathcal{C}\,.
\end{align}
This relation leads to:
\begin{align}
p_\parallel^2={}&\frac{1}{2\mathcal{A}^2\mathcal{C}^2}\Bigl[\mathcal{B}^2\mathcal{C}^2+\mathcal{A}^2\mathcal{D}^2+4nh\mathcal{A}\mathcal{C}\mathcal{E}^2\notag\\
&\pm (\mathcal{A}\mathcal{D}-\mathcal{B}\mathcal{C})\sqrt{(\mathcal{B}\mathcal{C}+\mathcal{A}\mathcal{D})^2+8nh\mathcal{A}\mathcal{C}\mathcal{E}^2}\Bigl]\,.
\end{align}
Evidently, the condition $\mathcal{A}\mathcal{D}\neq\mathcal{B}\mathcal{C}$ induced by the magnetic field, disrupts the energy degeneracy of spin states within the same Landau level. In contrast, in the absence of a magnetic field, where $\mathcal{A}\mathcal{D}=\mathcal{B}\mathcal{C}$, the Zeeman splitting vanishes.

The dispersion relation can also be reformulated in terms of $\mathcal{S}$, $\mathcal{V}_\parallel$, $\mathcal{V}_\perp$, $\mathcal{A}_v$ and $\mathcal{T}$ as follows:
\begin{align}
    \Delta ={}&-4p_\parallel^2(h\mathcal{S}\mathcal{A}_v -2h\mathcal{V}_\parallel \mathcal{T})^2\notag\\
    &+\left(\mathcal{S}^2-4h^2\mathcal{T}^2+h^2p^2_\parallel\mathcal{A}_v^2-p_\parallel^2\mathcal{V}_\parallel^2+2nh\mathcal{V}_\perp^2\right)^2=0\,.
\end{align}
This leads to:
\begin{align}
    p_\parallel^2=&\frac{\Omega_1   +\Omega_2-\Omega_3\pm2h\left(\mathcal{A}_v\mathcal{S}-2\mathcal{V}_\parallel\mathcal{T}\right)\sqrt{\Omega_4}}{\left(\mathcal{V}_\parallel^2-h^2\mathcal{A}_v^2\right)^2}\,,
\end{align}
where:
\begin{align}
\Omega_1&=h^2\mathcal{A}_v^2\left(\mathcal{S}^2+4h^2\mathcal{T}^2-2nh \mathcal{V}_\perp^2\right)\,,\notag\\
\Omega_2&=\mathcal{V}_\parallel^2\left(\mathcal{S}^2+4h^2\mathcal{T}^2+2nh\mathcal{V}_\perp^2\right)\,,\notag\\
\Omega_3&=8h^2\mathcal{S}\mathcal{V}_\parallel\mathcal{A}_v\mathcal{T}\,,\notag\\
\Omega_4&=(\mathcal{S}\mathcal{V}_\parallel-2h^2\mathcal{A}_v\mathcal{T})^2
    +2nh\mathcal{V}_\perp^2(\mathcal{V}_\parallel^2-h^2\mathcal{A}_v^2)\,.
\end{align}

\section{Dirac Structures of the Quark Propagator in a Magnetic Field}\label[appendix]{appendix:structure}

The quark propagator in a magnetic field can be expressed as a sum of Lorentz contractions involving tensor structures derived from the fundamental scalars, vectors, and tensors of the theory, along with the $16$ Dirac matrices.

In the presence of a magnetic field, the theory incorporates both the momentum $p_\mu$ and the electromagnetic field tensor $F_{\mu\nu}$. From these, four linearly independent invariants can be constructed. Among them, the purely field-independent invariants are Lorentz scalar $\mathcal{F}$ and pseudoscalar $\mathcal{G}$~\cite{Dittrich:2000zu}:
\begin{align}
	\mathcal{F}=\frac{1}{4}F_{\mu\nu}F^{\mu\nu}\,,\quad \mathcal{G}=\frac{1}{4}\tilde{F}_{\alpha\beta}F^{\alpha\beta}\,,
\end{align}
where $\tilde{F}_{\mu\nu}$ is the dual field strength tensor defined as:
\begin{align}
	\tilde{F}_{\mu\nu}=\frac{1}{2}\epsilon_{\mu\nu\alpha\beta}F^{\alpha\beta}\,,
\end{align}
with $\epsilon_{\alpha\beta\mu\nu}$ being the four-dimensional Levi-Civita symbol. The remaining two scalar invariants are:
\begin{align}
	z_p=(p_\alpha F^{\alpha\kappa})(p_\beta \tensor{F}{^\beta_\kappa})\,,\quad p^2=p^\mu p_\mu\,.
\end{align}

Additionally, four linearly independent vector and axial-vector combinations can be formed from $p_\mu$ and $F_{\mu\nu}$:
\begin{align}
	p^\mu\,,\quad F^{\mu\alpha} p_\alpha\,,\quad F^{\mu\alpha} F_{\alpha\beta}p^\beta \,,\quad \tilde{F}^{\mu\alpha}p_\alpha\,.
\end{align}
Higher powers of  $F_{\mu\nu}$ can be decomposed using fundamental algebraic relationships:
\begin{align}\label{Fmunurelation}
	&F^{\mu\alpha}\tensor{F}{^\nu_\alpha} - \tilde{F}^{\mu\alpha}\tensor{\tilde{F}}{^\nu_\alpha}=2\mathcal{F}g^{\mu\nu}\,,\notag\\
	&F^{\mu\alpha}\tensor{\tilde{F}}{^\nu_\alpha}=\tilde{F}^{\mu\alpha}\tensor{{F}}{^\nu_\alpha}=\mathcal{G}g^{\mu\nu}\,.
\end{align}

From these vectors and axial-vectors, one can construct second-rank tensors, resulting in ten distinct terms. Among these, three terms with more than two $F^{\mu\nu}$ can be expressed as functions of quadratic terms. Thus, we focus on seven terms involving fewer than two $F^{\mu\nu}$:
\begin{align}
	&p^\mu p^\nu \,,\quad
	F^{\mu\alpha} p_\alpha p^\nu\,,\quad
	\tilde{F}^{\mu\alpha}p_\alpha p^\nu \,,\quad
	F^{\mu\alpha}  F^{\nu\rho} p_\alpha p_\rho\,,\notag\\
	&\tilde{F}^{\mu\alpha} F^{\nu\rho} p_\alpha p_\rho\,,\quad
	F^{\mu\alpha} F_{\alpha\beta}p^\beta p^\nu	\,,\quad
	\tilde{F}^{\mu\alpha} \tilde{F}^{\nu\rho} p_\alpha p_\rho \,.
\end{align}
The last term is not linearly independent and can be expressed in terms of others:
\begin{align}
	\tilde{F}^{\mu\alpha} \tilde{F}^{\nu\rho} p_\alpha p_\rho
	=&\frac{1}{4}\epsilon^{\mu\alpha\kappa\gamma}F_{\kappa\gamma}\epsilon^{\nu\rho\lambda\delta}F_{\lambda\delta}p_\alpha p_\rho\,,
\end{align}
using properties of the product of two $\epsilon$ tensors: 
\begin{align}
	\epsilon^{\mu\alpha\kappa\gamma}\epsilon^{\nu\rho\lambda\delta}=\text{det}
	\begin{pmatrix}
		g^{\mu\nu}&g^{\mu\rho}&g^{\mu\lambda}&g^{\mu\delta}\\
		g^{\alpha\nu}&g^{\alpha\rho}&g^{\alpha\lambda}&g^{\alpha\delta}\\
		g^{\kappa\nu}&g^{\kappa\rho}&g^{\kappa\lambda}&g^{\kappa\delta}\\
		g^{\gamma\nu}&g^{\gamma\rho}&g^{\gamma\lambda}&g^{\gamma\delta}\\
	\end{pmatrix}\,.
\end{align}
Thus, we identify six linearly independent terms. Additionally, since the tensor is coupled to $\sigma^{\mu\nu}$, an antisymmetric second-rank tensor with six independent components, we can construct six linearly independent antisymmetric second-rank tensors. In addition to those mentioned above, $F^{\mu\nu}$ is also an antisymmetric tensor that couples directly with $\sigma_{\mu\nu}$. Furthermore, terms related to the energy-momentum tensor, such as $F^{\mu\rho}\tensor{F}{^\nu_\rho}$ and $g^{\mu\nu}F^{\rho\sigma}F_{\rho\sigma}$, can also be constructed.

In summary, using momentum $p_\mu$ and electromagnetic field strength $F_{\mu\nu}$, one can construct the following structures:
\begin{itemize}
	\item Three scalars: $\mathcal{F}$, $z_p$, and $p^2$.
	\item One pseudoscalar: $\mathcal{G}$.
	\item Three vectors: $p^\mu$, $q_fF^{\mu\alpha}p_\alpha$, and $q_f^2F^{\mu\alpha} F_{\alpha\beta}p^\beta$.
	\item One axial-vector: $q_f\tilde{F}^{\mu\alpha}p_\alpha$.
	\item Nine tensors:
	\begin{align}
	&q_fF^{\mu\nu}\,,\quad q_f^2F^{\mu\rho}\tensor{F}{^\nu_\rho}\,,\quad q_f^2g^{\mu\nu}F^{\rho\sigma}F_{\rho\sigma}\,,\notag\\
	&(g^{\mu\alpha}g^{\nu\rho}-g^{\nu\alpha}g^{\mu\rho} ) p_\alpha p_\rho \,,\notag\\
	&q_f(F^{\mu\alpha} g^{\nu\rho}-F^{\nu\alpha} g^{\mu\rho}) p_\alpha p_\rho \,,\notag\\
	&q_f(\tilde{F}^{\mu\alpha}g^{\nu\rho} -\tilde{F}^{\nu\alpha}g^{\mu\rho}) p_\alpha p_\rho \,,\notag\\	
	&q_f^2(F^{\mu\alpha}  F^{\nu\rho} -F^{\nu\alpha}  F^{\mu\rho}) p_\alpha p_\rho\,,\notag\\
	&q_f^2(\tilde{F}^{\mu\alpha} F^{\nu\rho} -\tilde{F}^{\nu\alpha} F^{\mu\rho}) p_\alpha p_\rho \,,\notag\\
	&q_f^2(F^{\mu\sigma} F_{\sigma\alpha}  g^{\nu\rho}-F^{\nu\sigma} F_{\sigma\alpha}  g^{\mu\rho}) p_\alpha p_\rho\,.
\end{align}
\end{itemize}
Note that for each $F_{\mu\nu}$, a factor of electric charge $q_f$ is included.

The possible contractions in theory must be invariant under charge conjugation (C), parity (P), and time-reversal (T) transformations. Focusing on charge conjugation, we have the following properties (particle to antiparticle):
\begin{align*}
    Cq_f C^{-1}&=-q_f\,,&
	C p^\mu C^{-1}&=-p^\mu\,,\notag\\
	C A^\mu C^{-1}&=-A^\mu \,,&
	C F^{\mu\nu} C^{-1}
		&=-F^{\mu\nu}\,,\notag\\
	C \tilde{F}^{\mu\nu} C^{-1}
	&= -\tilde{F}^{\mu\nu}\,,{\addtocounter{equation}{1}\tag{\theequation}}
\end{align*}
and for the $16$ Dirac matrices:
\begin{align*}
	C \mathds{1} C^{-1}&= \mathds{1}\,,&
	C \gamma_5 C^{-1}&= \gamma_5 \,,\\
	C \gamma_\mu C^{-1}&= -\gamma_\mu^T \,,&
	C \gamma_5\gamma_\mu C^{-1}&= (\gamma_5\gamma_\mu)^T \,,\\
	C \sigma_{\mu\nu} C^{-1}&= -\sigma_{\mu\nu}^T\,.{\addtocounter{equation}{1}\tag{\theequation}}
\end{align*}
Terms that do not comply with the following condition:
\begin{align}
C X C ^{-1} = X^T\,,
\end{align}
must be excluded from consideration, where $X$ represents the term under evaluation. Thus, the allowed terms are as follows:
\begin{itemize}
	\item Three scalars: $\mathcal{F}$, $z_p$, and $p^2$.
\item One pseudoscalar: $\mathcal{G}$.
	\item Two vectors: $p^\mu$ and $q_f^2F^{\mu\alpha} F_{\alpha\beta}p^\beta$.
	\item One axial-vector: $q_f\tilde{F}^{\mu\alpha}p_\alpha$.
	\item Three tensors: $q_fF^{\mu\nu}$, $q_f(F^{\mu\alpha} g^{\nu\rho}-F^{\nu\alpha} g^{\mu\rho}) p_\alpha p_\rho$, and $q_f(\tilde{F}^{\mu\alpha}g^{\nu\rho} -\tilde{F}^{\nu\alpha}g^{\mu\rho}) p_\alpha p_\rho $.
\end{itemize}
Notably, the last tensor term, $\tilde{F}^{\mu\alpha}g^{\nu\rho} p_\alpha p_\rho\sigma_{\mu\nu}$, is CP-odd and thus vanishes.

We can now outline the following invariants:
\begin{align}
	\mathbb{S}&=\mathcal{S}\,,\notag\\
	\mathbb{P}&=\mathcal{P}\mathcal{G}\,,\notag\\
	\mathbb{V}^\mu &=\mathcal{V}_1 p^\mu +\mathcal{V}_2 q_f^2 F^{\mu\alpha} F_{\alpha\beta}p^\beta\,,\notag\\
	\mathbb{A}^\mu &=\mathcal{A}q_f\tilde{F}^{\mu\alpha}p_\alpha\,,\notag\\
	\mathbb{T}^{\mu\nu}&=\mathcal{T}_1 q_f F^{\mu\nu}
	+\mathcal{T}_2q_f(F^{\mu\alpha} p^\nu-F^{\nu\alpha} p^\mu) p_\alpha\,,
\end{align}
where $\mathcal{S},\mathcal{P},\mathcal{A},\mathcal{V}_1,\mathcal{V}_2,\mathcal{T}_1,\mathcal{T}_2$ are scalar functions dependent on momentum and electromagnetic fields. The coupling of these invariants with the $16$ Dirac matrices yields the general structure of the inverse quark propagator: 
\begin{align}
	S^{-1}(p)= \mathbb{S}  + \mathbb{P} \gamma_5 + \mathbb{V}^\mu \gamma_\mu + \mathbb{A}^\mu \gamma_5\gamma_\mu +\mathbb{T}^{\mu\nu} \sigma_{\mu\nu}\,.
\end{align}

For a constant magnetic field along the $z$-axis, the nonzero components of $F_{\mu\nu}$ are $F_{12}=-F_{21}=-B$ and $\tilde{F}^{03}=-\tilde{F}^{30}=-B$. Thus, we have:
\begin{align}
	\mathcal{G}&=0\,,\notag\\
	q_f^2 F^{\mu\alpha} F_{\alpha\beta}p^\beta \gamma_\mu 
	&=-h^2 \slashed{p}_\perp\,,\notag\\
	q_f\tilde{F}^{\mu\alpha}p_\alpha \gamma_5\gamma_\mu
	&=h \Sigma^3\slashed{p}_\parallel\,,\notag\\
	q_f F^{\mu\nu}\sigma_{\mu\nu}
	&=2h\Sigma^{3}\,,\notag\\
	q_f(F^{\mu\alpha} p^\nu-F^{\nu\alpha} p^\mu) p_\alpha\sigma_{\mu\nu}
	&=-2hp_\perp^2 \Sigma^{3}\,.
\end{align}
Consequently, the inverse quark propagator can be expressed as: 
\begin{align}
	S^{-1}(p)
	=\mathcal{S}   + \mathcal{V}_\parallel \slashed{p}_\parallel + \mathcal{V}_\perp \slashed{p}_\perp +\mathcal{A} h \Sigma_3\slashed{p}_\parallel +2h \mathcal{T}\Sigma_3\,,
\end{align}
where the following definitions apply:
\begin{align*}
	\gamma_5\gamma_0&=\Sigma^3\gamma_3\,,&
	\gamma_5\gamma_3&=\Sigma^3\gamma_0\notag\\
	\mathcal{V}_\parallel &= \mathcal{V}_1\,,&
	\mathcal{V}_\perp&= \mathcal{V}_1-\mathcal{V}_2 h^2\,,\notag\\
	\mathcal{T}&=\mathcal{T}_1-\mathcal{T}_2p_\perp^2\,.{\addtocounter{equation}{1}\tag{\theequation}}
\end{align*}
Note that there are five unspecified scalar functions: $\mathcal{S},\mathcal{V}_\parallel,\mathcal{V}_\perp,\mathcal{A},\mathcal{T}$. This general form aligns precisely with the results found in Refs.~\cite{Ritus:1972ky,Ferrer:1998vw}.

\section{Explicit Form of the Gap Equation in Euclidean Space}\label[appendix]{appendix:gapeq}

The structures of the inverse propagator are given in Eq.~\eqref{eq:inversestructures}, and those of the propagator are given in Eq.~\eqref{eq:prostructures}. By substituting these into the gap equation, Eq.~\eqref{eq:gapeqmom}, and projecting onto various Dirac components, we derive a set of scalar equations. To facilitate numerical computation, we perform a Wick rotation to Euclidean space, where the dressing functions depend on longitudinal momentum $p_l^2=p_0^2+p_z^2$ and transverse momentum $p_t^2=p_x^2+p_y^2$. The resulting equations are explicitly expressed as follows (with $q=k-p$):
\begin{align}\label{EqsAtoE}
	\hat{A}&=Z_2-Z_2^2g^2C_F  \int\frac{d^4{k}}{(2\pi)^4}\frac{G(q^2)}{q^2}\frac{\mathbb{K}_A}{\Delta_E}\,,\notag\\
	\hat{B}&=Z_2Z_m-Z_2^2g^2C_F  \int\frac{d^4{k}}{(2\pi)^4}\frac{G(q^2)}{q^2}\frac{\mathbb{K}_B}{\Delta_E}\,,\notag\\
	\hat{C}&=Z_2-Z_2g^2C_F  \int\frac{d^4{k}}{(2\pi)^4}\frac{G(q^2)}{q^2}\frac{\mathbb{K}_C}{\Delta_E}\,,\notag\\
	\hat{D}&=Z_2Z_m-Z_2^2g^2C_F  \int\frac{d^4{k}}{(2\pi)^4}\frac{G(q^2)}{q^2}\frac{\mathbb{K}_D}{\Delta_E}\,,\notag\\
	\hat{E}&=Z_2-Z_2^2g^2C_F  \int\frac{d^4{k}}{(2\pi)^4}\frac{G(q^2)}{q^2}\frac{\mathbb{K}_E}{\Delta_E}\,.
\end{align}
The kernels are defined as:
\begin{align}
	\mathbb{K}_A=&N_A K_{AA}+N_C K_{AC} -N_E K_{AE}\,,\notag\\
	\mathbb{K}_B=&N_B K_{BB}+N_D K_{BD}\,,\notag\\
	\mathbb{K}_C=&N_A K_{CA}+N_C K_{CC} -N_E K_{CE}\,,\notag\\
	\mathbb{K}_D=&N_B K_{DB}+N_D K_{DD}\,,\notag\\
	\mathbb{K}_E=&N_A K_{EA}+N_C K_{EC} -N_E K_{EE}\,.
\end{align} 
The denominator is defined as:
\begin{align}
	\Delta_E={k}_l^2\hat{A}\hat{C} + k_t^2 \hat{E}^2+\hat{B}\hat{D}\,.
\end{align}
The terms ${N}_A$ to ${N}_E$ are given by:
\begin{align}
{N}_A=&\hat{C} - \frac{h\hat{E}^2 \hat{C}}{\Delta_E} + \frac{ \hat{D}(\hat{A}\hat{D}-\hat{B}\hat{C})}{\Delta_E}\,,\notag\\
	{N}_B=&\hat{D} - \frac{h\hat{E}^2 \hat{D}}{\Delta_E}  - \frac{{k}_l^2 \hat{C}(\hat{A}\hat{D}-\hat{B}\hat{C})}{\Delta_E}  \,,\notag\\
	{N}_C=&\hat{A} + \frac{h\hat{E}^2 \hat{A}}{\Delta_E} - \frac{\hat{B}(\hat{A}\hat{D}-\hat{B}\hat{C})}{\Delta_E} \,,\notag\\
	{N}_D=&\hat{B} + \frac{h\hat{E}^2 \hat{B}}{\Delta_E}  +\frac{{k}_l^2 \hat{A}(\hat{A}\hat{D}-\hat{B}\hat{C})}{\Delta_E}\,,\notag\\
	{N}_E=&\hat{E}\,.
\end{align}
The $K$ matrices are defined as:
\begin{align}
	 K_{AA}=&\frac{1}{q^2{p}_l^2}\left[{q}_l^2({p}_l\cdot{k}_l)-2({p}_l\cdot {q}_l)({k}_l\cdot {q}_l)\right]\,,\notag\\
	 K_{AC}=&\frac{1}{q^2{p}_l^2}\left[-q^2({p}_l\cdot{k}_l)-{q}_l^2 ({p}_l\cdot{k}_l)\right]\,,\notag\\
	 K_{AE}=&\frac{1}{q^2{p}_l^2}[2({p}_l\cdot {q}_l)(k_t\cdot q_t)]\,,\notag\\
	 K_{BB}=&\frac{{q}_l^2}{q^2}-2\,,\quad
	 K_{BD}=-1-\frac{{q}_l^2}{q^2}\,,\notag\\
	 K_{CA}=&K_{AC}\,,\quad
	 K_{CC}=K_{AA}\,,\quad
	 K_{CE}=K_{AE}\,,\notag\\
	 K_{DB}=&K_{BD}\,,\quad
	 K_{DD}=K_{BB}\,,\notag\\
	 K_{EA}=&-\frac{1}{q^2p_t^2}\left[({k}_l\cdot {q}_l)(p_t\cdot q_t)\right]\,,\notag\\
	 K_{EC}=&K_{EA}\,,\notag\\
	 K_{EE}=&\frac{(p_t\cdot k_t)}{p_t^2}+ 2\frac{(p_t\cdot q_t)(k_t\cdot q_t)}{q^2p_t^2}\,.
\end{align}
Solving Eq.~\eqref{EqsAtoE} yields solutions for $\hat{A}-\hat{E}$ (now functions of $p_l^2$ and $p_t^2$), allowing us to determine the quark propagator.




\bibliography{bibreferences}

\end{document}